\documentclass{aa}  

\usepackage[usenames, dvipsnames]{xcolor}
\usepackage{graphicx}
\usepackage{txfonts}
\usepackage{amsmath}
\usepackage{textgreek}
\usepackage{natbib,twoopt}
\usepackage[colorlinks=true,allcolors=blue]{hyperref} 
\bibpunct{(}{)}{;}{a}{}{,} 
\makeatletter
\newcommandtwoopt{\citeads}[3][][]{\href{http://adsabs.harvard.edu/abs/#3}%
{\def\hyper@linkstart##1##2{}%
\let\hyper@linkend\@empty\citealp[#1][#2]{#3}}}
\newcommandtwoopt{\citepads}[3][][]{\href{http://adsabs.harvard.edu/abs/#3}%
{\def\hyper@linkstart##1##2{}%
\let\hyper@linkend\@empty\citep[#1][#2]{#3}}}
\newcommandtwoopt{\citetads}[3][][]{\href{http://adsabs.harvard.edu/abs/#3}%
{\def\hyper@linkstart##1##2{}%
\let\hyper@linkend\@empty\citet[#1][#2]{#3}}}
\newcommandtwoopt{\citeyearads}[3][][]%
{\href{http://adsabs.harvard.edu/abs/#3}
{\def\hyper@linkstart##1##2{}%
\let\hyper@linkend\@empty\citeyear[#1][#2]{#3}}}
\makeatother
\usepackage{siunitx}
\DeclareSIUnit \parsec {pc}
\DeclareSIUnit\angstrom{\text {Å}}
\DeclareSIUnit\year{\text {yr}}
\DeclareSIUnit\erg{\text {erg}}
\DeclareSIUnit\jansky{\text {Jy}}
\DeclareSIUnit \solarmass {\ensuremath{M_\odot}}
\DeclareSIUnit \h {\ensuremath{\mathit{h}}}

\usepackage{ulem}

\usepackage{comment}

\newcommand{\fedd}{f_\mathrm{Edd}}
\newcommand{\Msun}{M_\odot}

\begin{document} 

\title{Multi-messenger study of merging massive black holes in the \textsc{Obelisk} simulation: Gravitational waves, electromagnetic counterparts, and their link to galaxy and black-hole populations}
   \titlerunning{Multi-messenger study of massive black hole mergers in the \textsc{Obelisk} simulation}


   \author{Chi An Dong-P{\'a}ez\inst{1}
          \and
          Marta Volonteri\inst{1}
          \and
          Ricarda S. Beckmann\inst{2}
          \and
          Yohan Dubois\inst{1}
          \and
          Alberto Mangiagli\inst{3}
          \and
          Maxime Trebitsch\inst{4}
          \and
          Susanna D. Vergani\inst{5,1}
          \and
          Natalie A. Webb\inst{6}
          }

   \institute{Institut d’Astrophysique de Paris, UMR 7095, 
                CNRS and Sorbonne Universit\'{e}, 98 bis boulevard Arago, 75014 Paris, France\\
              \email{dongpaez@iap.fr}
         \and
              Institute of Astronomy and Kavli Institute for Cosmology, University of Cambridge, Madingley Road, Cambridge, CB3 0HA, UK
         \and
             AstroParticule et Cosmologie, Universit{\'e} Paris, CNRS, Astroparticule et Cosmologie, 75013 Paris, France
         \and
             Kapteyn Astronomical Institute, University of Groningen, P.O. Box 800, 9700 AV Groningen, The Netherlands
         \and
             GEPI, Observatoire de Paris, Université PSL, CNRS, 5 Place Jules Janssen, 92190 Meudon, France
         \and
             CNRS, IRAP, 9 Av. colonel Roche, BP 44346, 31028 Toulouse cedex 4, France
        }
        
   \date{Received ; accepted }

 
  \abstract{
  Massive black-hole (BH) mergers are predicted to be powerful sources of low-frequency gravitational waves (GWs). Coupling the detection of GWs with an electromagnetic (EM) detection can provide key information about merging BHs and their environments as well as cosmology. We study the high-resolution cosmological radiation-hydrodynamics simulation \textsc{Obelisk}, run to redshift $z=3.5$, to assess the GW and EM detectability of high-redshift BH mergers, modelling spectral energy distribution and obscuration. For EM detectability, we further consider sub-grid dynamical delays in postprocessing.
  We find that most of the merger events can be detected by LISA, except for high-mass mergers with very unequal mass ratios. Intrinsic binary parameters are accurately measured, but the sky localisation is poor generally. Only $\sim 40\%$ of these high-redshift sources have a sky localisation better than $10\,\mathrm{deg}^2$. Merging BHs are hard to detect in the restframe UV since they are fainter than the host galaxies, which at high redshift are star-forming. A significant fraction, $15$ to $35\%$, of BH mergers instead outshine the galaxy in X-rays, and about $5-15\%$ are sufficiently bright to be detected with sensitive X-ray instruments. If mergers induce an Eddington-limited brightening, up to $30\%$ of sources can become observable. The transient flux change originating from such a brightening is often large, allowing $4-20\%$ of mergers to be detected as EM counterparts. 
  A fraction, $1-30\%$, of mergers are also detectable at radio frequencies. Transients are found to be weaker for radio-observable mergers. Observable merging BHs tend to have higher accretion rates and masses and are overmassive at a fixed galaxy mass with respect to the full population. Most EM-observable mergers can also be GW-detected with LISA, but their sky localisation is generally poorer. This has to be considered when using EM counterparts to obtain information about the properties of merging BHs and their environment. 
  }

   \maketitle
%

\section{Introduction}

The merger of two neutron stars detected as both a gravitational wave (GW) and electromagnetic (EM) source \citep{2017ApJ...848L..12A} has recently opened up the field of multi-messenger studies of astrophysical phenomena. Another promising candidate for such multi-messenger studies is the merger of two massive black holes (BHs). Merging BHs with masses $\sim 10^4-10^7\,\Msun$ can be detected in GWs by the future space-based interferometer LISA  \citep[Laser Interferometer Space Antenna,][]{2023LRR....26....2A} as well as by  the proposed missions TianQin \citep{2016CQGra..33c5010L} and Taiji \citep{2020IJMPA..3550075R}. The horizon of these detectors is large, with the ability to detect merging BHs out to $z\sim 10$ for mass ratios not too far from unity. Mergers of such massive BHs are also expected to be detectable electromagnetically, as massive BHs are generically surrounded by gas in galactic centres and they are therefore associated with luminous sources such as active galactic nuclei (AGN) when they accrete such gas. If these merging BHs can also be detected electromagnetically, then we can use them to study accretion physics in dynamical spacetimes, to obtain independent measures of BH masses  
\citep{2023LRR....26....2A}, and to constrain fundamental physics \citep{2022LRR....25....4A} and cosmology \citep{2022arXiv220405434A}.

While massive BHs have been detected electromagnetically for many years, in the form of AGN, it remains very unclear whether they give off a sufficiently distinguishable signal at the moment of merger to allow for a multi-messenger study. Over the years, many models and simulations have been developed for the actual physics of the production of an EM counterpart at the merger \citep[e.g.][]{Armitage2002,2008ApJ...684..835S,2010MNRAS.401.2021R,2012MNRAS.420..860S,2014ApJ...785..115R,2022ApJ...928..137G,Kelly2021}. 

Numerical studies have shown that a change in the luminosity occurs around the time of the merger for BH binaries evolving in circumbinary discs. Before merging, during the late inspiral of gas-rich BH binaries, the binary torques excavate a low-density cavity in the circumbinary disc. Consequently, the circumbinary disc acquires an inner rim at a radius of the order of the BH binary semi-major axis. Material pile sup at this inner rim, creating a high-density region or a non-axisymmetric `lump' \citep[e.g.][]{Kocsis2012,Noble2012}. Despite the potential barrier that maintains the cavity, gas streams can flow through it and feed accretion minidiscs around the individual BHs \citep[e.g.][]{Noble2012,Shi2015,Tang2018}. At some point during the binary evolution, the rate at which the orbit shrinks due to the emission of gravitational waves is faster than the viscous timescale -- the disc cannot evolve fast enough and decouples from the binary \citep[e.g.][]{Milosavljevic2005}. Although the gas accretion streams can continue to feed the BHs virtually until the merger, the accretion rate can decrease after the binary completely decouples from the disc \citep[e.g.][]{Gold2014b,Farris2015}. The minidiscs, which dominate the hard X-ray emission \citep[e.g.][]{dAscoli2018}, could also gradually disappear near the merger, causing a drop in the X-ray luminosity \citep{Tang2018}. In contrast to this, \cite{Armitage2002} and \cite{2016MNRAS.457..939C} suggest that squeezing and rapid accretion in the minidisc of the primary BH could lead to large enhancements in the accretion rate and luminosity. 

After the BH merger, the disc is expected to maintain initially a central cavity at $\sim 10-20$ gravitational radii, causing a small diminution (less than a factor of $\sim 3$ for a non-spinning BH) in the radiative efficiency compared to that of a disc around a single BH \citep{Bogdanovic2022}. The cavity is then refilled on a timescale ${t_\mathrm{vis} \sim 0.1 (M_\bullet/10^6\,M_\odot)(\alpha/0.1)^{-8/5}(h/0.1)^{-16/5}}\,\mathrm{yr}$, where $\alpha$ is the disc viscosity parameter and $h$ is the disc aspect ratio \citep[e.g.][]{Milosavljevic2005,Farris2015,Yuan2021}. The disc refilling gradually increases the BH accretion rate and luminosity. Moreover, when the pileup of material at the inner rim is accreted, the larger availability of gas can potentially drive a post-merger luminosity burst.

Changes in jet properties have also been suggested in conjunction with BH mergers \citep{2002Sci...297.1310M}. The merger can modify BH properties, such as $M_\bullet$, $f_\mathrm{Edd}$, or $a$, or the properties of gas dynamics and magnetic fields around the remnant BH, which can lead to observable radio signatures on shorter scales, of the order of hours or days.
This is supported by results from general relativistic magnetohydrodynamical simulations of BH mergers \citep{Palenzuela2010,Moesta2012,Gold2014b,Kelly2017,Kelly2021,Cattorini2021,Cattorini2022}. 
For example, simulations predict the production of a flare of Poynting luminosity at the merger, lasting for $\sim 0.1 \,\mathrm{day}\, (M_\bullet/10^6\, M_\odot)$. 
This Poynting luminosity can be comparable to that of the pre-merger jet, although it seems to depend on various parameters (mass ratio, spin magnitude and alignment, gas density, magnetic field, etc.) which have not been extensively explored in simulations. \citet{Yuan2021} modelled the jet spectrum under the assumption that a newly formed jet after the merger impacts and shocks the nearby gas, thus producing a source similar to a gamma-ray burst.  

Fewer studies have been applied to BH merger populations to estimate the number and properties of BH mergers with an EM counterpart \citep{2006MNRAS.372..869D,Tamanini2016,2019MNRAS.485.1579K,2019ApJ...879..110K,Mangiagli2022,2023MNRAS.519.5962L, Chakraborty2023}. 
In the following, we study the properties of the BH merger population in the \textsc{Obelisk} simulation \citep{Trebitsch2021} and assess the multi-messenger observability of their corresponding GW events and EM counterparts, as well as the biases of the observable population. \textsc{Obelisk} is a cosmological radiation-hydrodynamical simulation evolving a protocluster down to redshift $\sim 3.5$. This simulation is ideal for our purposes since it has a high resolution ($\sim 35\,\si{\parsec}$) and incorporates detailed models for a wide range of BH physical processes, such as accretion, feedback, spin evolution, and dynamical friction, which are key in order to produce a realistic BH merger population. We remind the reader that \textsc{Obelisk} models the evolution of an overdense region, and thus it cannot be used to predict merger rates in an unbiased way. This work follows on from \citet{Dong-Paez2023a} in which we present and analyse the population of BH mergers in comparison to the total population of BH in \textsc{Obelisk}.

In Section~\ref{sec:Method}, we summarise the properties of the \textsc{Obelisk} simulation and the identification and selection criteria of galaxies and BH mergers. We also describe our calculation of sub-grid merger delays, BH luminosities in several EM bands, and our simulations of the GW parameter estimation by LISA. We present our results in the subsequent sections -- in Section~\ref{subsec:GW_observability}, we study the GW observability and parameter estimation by LISA of the BH merger sample, and in Sections~\ref{subsec:UV_X-rays} and~\ref{subsec:radio} we study their observability in several EM bands (X-rays, UV, and radio), the bias in the properties of the observable mergers with respect to the unobserved sample and the synergies with the GW detections. In Section~\ref{sec:comparison}, we discuss our methods and results in the context of the previous work. Finally, in Section~\ref{sec:conclusions}, we conclude and summarise our main results.

\section{Method}
\label{sec:Method}

\subsection{The \textsc{Obelisk} simulation}
\label{subsec:Obelisk}

\begin{figure*}
    \includegraphics[width=\textwidth]{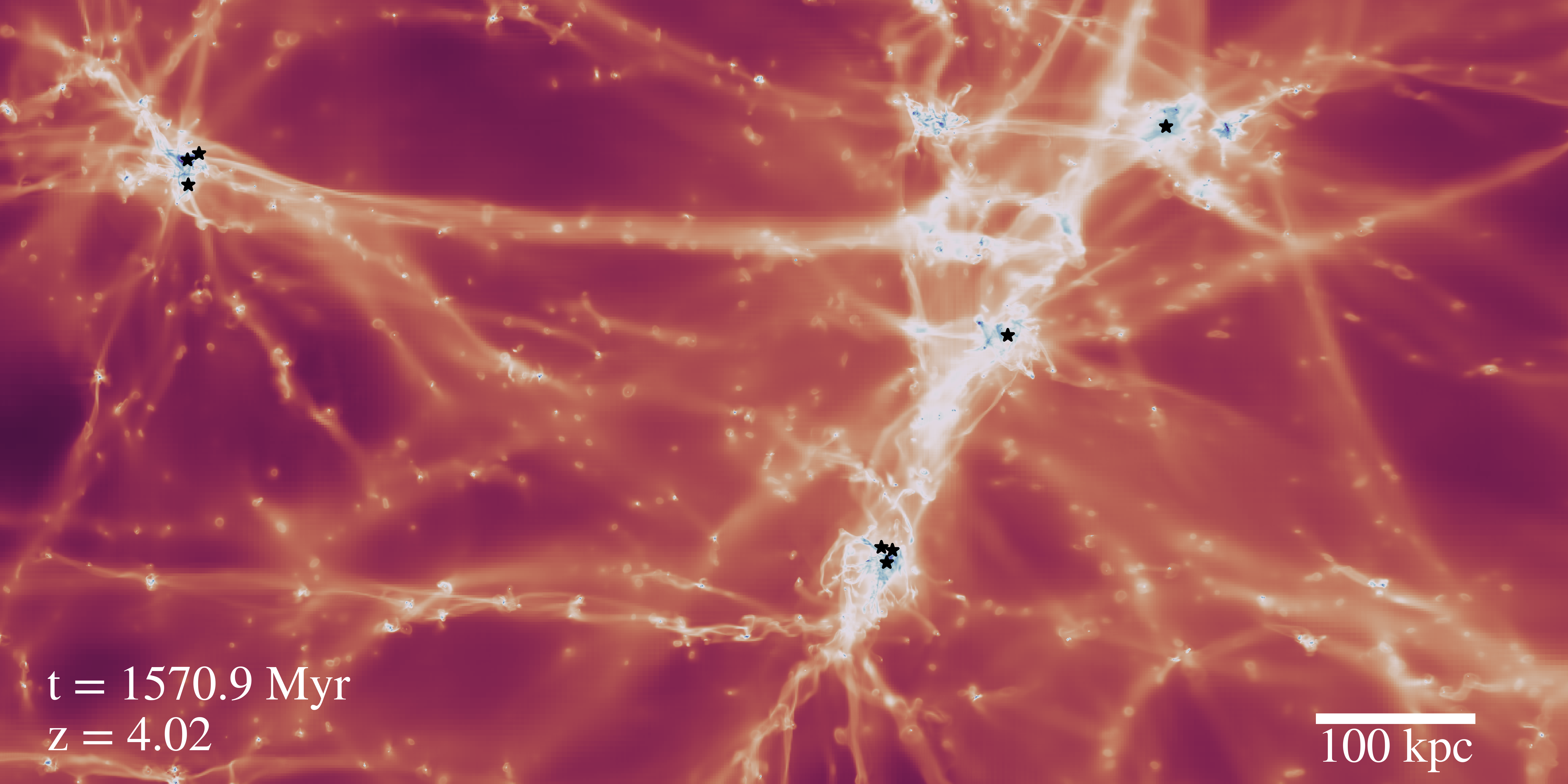}
    \caption{Gas density projection of a region of the \textsc{Obelisk} simulation at redshift $4.02$. BHs with $M_\bullet>10^6\,\Msun$ are denoted by black star symbols.}
    \label{fig:Obelisk_snap}
\end{figure*}

\textsc{Obelisk} \citep{Trebitsch2021} re-simulates at high-resolution ($\Delta x \simeq \SI{35}{\parsec}$) the most massive halo in \textsc{Horizon-AGN} at $z=1.97$ in the \textsc{Horizon-AGN} \citep{Dubois2014c} volume until redshift $z \simeq 3.5$. In Fig.~\ref{fig:Obelisk_snap}, we show the projected gas density in a region of the simulation at $z\sim 4$. Below we present a brief summary of the properties of the simulation. For a more detailed description, we refer the reader to \citet{Trebitsch2021} and \citet{Dong-Paez2023a}.

The simulation assumes a \textLambda CDM cosmology with  WMAP-7 parameters \citep{Komatsu2011} -- Hubble constant $H_0 = \SI{70.4}{\kilo\meter\per\second\per\mega\parsec}$, dark energy density parameter $\Omega_\Lambda = 0.728$, total matter density parameter $\Omega_\mathrm{m} = 0.272$, baryon density parameter $\Omega_\mathrm{b} = 0.0455$, amplitude of the power spectrum $\sigma_8=0.81$, and spectral index $n_\mathrm{s}=0.967$. The zoomed-in region, with a volume of $\sim 10^4 \, h^{-3}\,\mathrm{cMpc}^3$, was simulated with a DM mass resolution of $1.2\times 10^6\, \Msun$, while the remaining volume of the original $100 \, h^{-1} \, \mathrm{cMpc}$ \textsc{Horizon-AGN} box maintained a lower resolution. 

 \textsc{Obelisk} was run with \textsc{Ramses-RT} \citep{Rosdahl2013,Rosdahl2015}, a radiative transfer hydrodynamical code which builds on the adaptive mesh refinement (AMR) \textsc{Ramses} code \citep{Teyssier2002}. Cells are refined up to a smallest size of $35\,\si{\parsec}$ if its mass exceeds $8$ times the mass resolution. The simulation assumes an ideal monoatomic gas with adiabatic index $\gamma = 5/3$ and includes gas cooling and heating down to very low temperatures ($50\,\rm K$) with non-equilibrium thermo-chemistry for hydrogen and helium, and contribution to cooling from metals (at equilibrium with a standard ultraviolet background) released by SNe. 
 
 Stellar particles have a mass of $\sim 10^4\,\Msun$, and assume a \citet{Kroupa2001} initial mass function between $0.1$ and $100 \,\Msun$. Stars form in gas cells with density higher than $5\,\mathrm{H}\,\si{\per\cm\cubed}$ and Mach number $\mathcal{M}\geq 2$. The star formation efficiency depends on the local gas density, sound speed, and turbulent velocity. SN feedback takes place 
  $5\,\si{\mega\year}$ after the birth of a stellar particle, with a mass fraction of $0.2$, and releasing $10^{51}\,\si{\erg}$ per SN.  \textsc{Obelisk} also includes modelling of dust as a passive variable. 

BHs form when in a given cell both gas and stars exceed a density threshold of $100 \, \si{H} \, \si{\per\centi\meter\cubed}$ and the gas is Jeans unstable. The initial mass is  $M_{\bullet,0} =3\times 10^4\,\Msun$ and an exclusion radius of 50 comoving kpc is enforced to avoid formation of multiple BHs. 
Gas accretion is modelled using Bondi-Hoyle-Lyttleton (BHL) formalism,
\begin{equation}
    \dot{M}_\mathrm{BHL} = \frac{4\pi G^2 M_\bullet^2 \bar{\rho}}{\left(\bar{c_s}^2 + \bar{v}_\mathrm{rel}^2\right)^{3/2}},
    \label{eq:BHL_accretion}
\end{equation}
where $\bar{\rho}$, $\bar{c_s}$ and $\bar{v}_\mathrm{rel}$ are the local average gas density, gas sound speed, and BH relative velocity with respect to the gas. The accretion rate is capped at the Eddington rate
\begin{equation}
    \dot{M}_\mathrm{Edd} = \frac{4\pi GM_\bullet m_p}{\varepsilon_r \sigma_\mathrm{T} c},
\end{equation}
where $m_p$ is the proton mass, $\varepsilon_r$ is the radiative efficiency, $\sigma_\mathrm{T}$ is the Thompson cross-section, and $c$ is the speed of light. A fraction $\varepsilon_r$ of the accretion power $\dot{M}c^2$ is radiated, while the remaining $1-\varepsilon_r$ is accreted onto the BH, contributing to its mass growth. 
BHs with an Eddington ratio as $f_\mathrm{Edd} = \dot{M}/\dot{M}_\mathrm{Edd}<0.01$ (here $\dot{M} = \min{\{\dot{M}_\mathrm{BHL},\dot{M}_\mathrm{Edd}\}}$), are assumed to be radiatively inefficient, and radiative efficiency is reduced by a factor $f_\mathrm{Edd}/0.01$.

AGN feedback is modelled with a dual-mode approach. At $f_\mathrm{Edd}<0.01$, the AGN releases a fraction $\varepsilon_\mathrm{MCAF}$ of the rest-mass accreted energy as kinetic energy in jets. Jets assume Magnetically Chocked Accretion Flows, and $\varepsilon_\mathrm{MCAF}$ is a polynomial fit to the simulations of \citet{McKinney2012} as a function of BH spin. At higher $f_\mathrm{Edd}\geq 0.01$, the $15\%$ of the accretion luminosity is released isotropically  as thermal energy. 

Two BHs are merged when their separation becomes smaller than $4\Delta x$ and they are gravitationally bound. The simulation models dynamical friction explicitly, including both gas and collisionless particles (stars and DM) \citep[following the implementation presented in][]{Dubois2013,Pfister2019}. 

BH spins are self-consistently evolved on the fly via gas accretion and BH-BH mergers following \cite{Dubois2014a}. The model for BHs with $f_\mathrm{Edd}>0.01$ includes evolution of both spin magnitude \citep{Bardeen1970} and direction \citep{King05}, while for $f_\mathrm{Edd}<0.01$) rotational energy is assumed to power the radio jets and therefore the magnitude of BH spins can only decrease. We adopt the polynomial fits in \citet{McKinney2012}, with the same procedure for the update of the spin direction as for the $f_\mathrm{Edd}\geq 0.01$ case. Spin also evolves following the coalescence of two BHs using an analytical fit from \citet{Rezzolla2008a}. The value of the spin is used to determine the efficiency of the energy injection into jets in the radio mode, and the BH radiative efficiency.

\subsection{Galaxy catalogues and BH-galaxy matching}

The galaxy and BH merger catalogues used here are identical to those presented in \cite{Dong-Paez2023a}. We summarise the relevant procedure here and in the next section, but refer the reader to that paper for further details.

Galaxies and their DM haloes were identified together, using a version of the \textsc{AdaptaHOP} algorithm \citep{Aubert2004, Tweed2009} designed to work on both stars and DM particles. Substructures were identified using the most massive sub-maximum method, with a minimum number of particles (stars + DM) of 100 \citep[see details in][]{Trebitsch2021}. As \textsc{Obelisk} is a zoom simulation, we only considered halos that do not contain any low-resolution DM particle to avoid artefacts. In simulated high-redshift galaxies, disturbed morphologies are common, which makes it challenging to define the centre of a galaxy.  We followed what has been done for the \textsc{New-Horizon} simulation \citep{Dubois2021} and chose as our fiducial `centre' the position of the density peak (for stars, DM, and both) determined recursively using a shrinking sphere approach.

Since BHs are not artificially pinned to galaxy centres, we have to assign BHs to galaxies. The main BH of a galaxy was defined as the most massive BH located within $\max(4\Delta x,r_{50})$, where $r_{50}$ is the half-mass radius $r_{50}$. BHs that are not assigned to any galaxy as main BHs can be assigned as satellite BHs to the highest stellar mass galaxy enclosing them within $2r_{50}$. Finally, star formation rates were averaged over $\SI{10}{\mega\year}$. We note that galaxy properties were stored in snapshots, recorded every 15~Myr or less. BH properties were instead recorded at every coarse timestep, about 0.1~Myr. 

\subsection{Selection of BH mergers}
\label{subsec:merger_selection}

BHs that are merged in the simulation following the sub-grid algorithm for BH-BH mergers (at a distance of $4 \Delta x$) were identified as `numerical mergers'. To find the mergers host galaxy, we identified in the snapshot immediately after the merger which galaxy contains the location of the merger within a distance $<r_{50}$ from the galaxy centre. BH mergers occurring at larger distances from all galaxy centres were considered spurious cases and discarded. To account for the continued dynamical decay of the BH binary below $4 \Delta x$ we calculated delays in post-processing, and defined `delayed mergers' as  the outcome of adding such delays. For numerical mergers, the BH properties are measured at the coarse timestep immediately preceding, and the galaxy properties at the simulation snapshot prior to the merger. For numerical and delayed remnants, we measured galaxy properties at the first available post-merger output and BH properties at the first available post-merger coarse timestep.

Sub-grid merger delays were modelled as in \cite{Volonteri2020} and \citet{Dong-Paez2023a}. We included a dynamical friction phase from the position where the BHs were located at the numerical merger down to the centre of the host galaxy by computing the dynamical friction timescale for a massive object in an isothermal sphere, considering only the stellar component of the galaxy and including a factor $0.3$ to account for typical orbits being non-circular. We calculated the sinking time of both BHs in the numerical merger and took the longest of the two. For binaries whose dynamical friction timescale ends before the simulation is stopped at $z=3.5$ we further calculated binary evolution timescales through interaction with stars \cite{2015MNRAS.454L..66S} and gas \cite{2015MNRAS.448.3603D} until gravitational waves take over. Delayed mergers predicted to occur at $z<3.5$, after the final redshift down to which \textsc{Obelisk} has been run, cannot be modelled as we lack the information on host galaxy properties required to compute the relevant timescales.

We note that during this post-processed dynamical evolution, the BHs have already been merged numerically in the simulation. That is, the simulation does not track the individual evolution of the two BHs during the merger delays, but only that of a numerically merged BH with the total mass. Consequently, for delayed mergers, merger parameters that require individual properties of the BHs, such as the mass ratio or the pre-merger BH spins, cannot be extracted from the simulation. These parameters would need to be estimated in post-processing, and the final value would be strongly dependent on the model used \citep[e.g.][]{Farris2014,Duffell2020,Munoz2020,Siwek2020}. Since mass ratio is a key parameter for GW analysis, we excluded delayed mergers from any analysis involving the mass ratios or spins at merger. We considered both numerical mergers and delayed mergers as a way to bracket our uncertainty.

\subsection{AGN spectral energy distribution}
\label{subsec:SED}

Commonly, the AGN luminosity in different bands is estimated using bolometric corrections \citep[e.g.][]{Shen2020}, which are derived from mean observed quasar Spectral Energy Distributions (SEDs). Due to observational selection effects, the quasar samples used to calibrate such models tend to cover only a reduced region of the BH parameter space. In contrast, our simulated BH sample spans a much wider range in $M_\bullet$ and  $f_\mathrm{Edd}$.

In order to capture qualitatively the effect of BH physical parameters for a large region of the parameter space, it is often preferable to model AGN emission by adopting physically motivated analytical models \citep[e.g.,][]{2018MNRAS.480.1247K}. These models should converge to the standard SEDs used to calculate bolometric corrections when they are restricted to the range of $M_\bullet$ and $f_\mathrm{Edd}$ that characterise the quasar samples used to calibrate the bolometric corrections \citep[see for instance the Appendix in][]{Volonteri2017}. We modelled the AGN SED as the sum of the emission from a self-gravitating relativistic, geometrically thin, optically thin accretion disc \citep{Novikov1973} and a power-law X-ray emission with an exponential cutoff from the corona,
\begin{equation}
    L_\nu=
    \begin{cases}
        \mathcal{A}L_{\nu,\mathrm{disc}} & \nu\leq\nu_\mathrm{max}\\
        
        \mathcal{A}L_{\nu,\mathrm{disc}} + \mathcal{B}L_{\nu,\mathrm{disc}}(c/\SI{2500}{\angstrom})\left(\frac{\nu}{(c/\SI{2500}{\angstrom})}\right)^{\alpha_\mathrm{X}}e^{-\nu/\nu_\mathrm{c}}
        & \nu>\nu_\mathrm{max} \\
    \end{cases}
    \label{eq:SED_model}
\end{equation}
where $\mathcal{A}$ and $\mathcal{B}$ are normalisation constants and $L_{\nu,\mathrm{disc}}$ is the emission from a Novikov-Thorne accretion disc. 
In Eq.~\ref{eq:SED_model}, the first term corresponds to the thermal disc contribution, and the second term corresponds to a power-law emission dominating at high energy. The second term is switched on at $\nu_\mathrm{max}$ the frequency at which the Novikov-Thorne solution peaks. We fix the power-law index of the second term to $\alpha_\mathrm{X} = -0.9$ and the characteristic cutoff frequency to $\nu_\mathrm{c}=\SI{300}{\kilo\electronvolt}/h$ \citep[e.g.][]{Shen2020}.

The disc is assumed to radiate as a blackbody at each radius, with a temperature given by the Novikov-Thorne solution \citep{Krolik1999}. 
The total disc SED was then obtained by integrating the blackbody emission of each annulus over the radial extent of the disc. The disc is assumed to extend from the radius of the innermost stable circular orbit $r_\mathrm{isco}$. We set the maximum radius of the disc where the disc becomes self-gravitating. For a radiation pressure-dominated disc with opacity given by electron scattering, this radius is given by \citep{Laor1989}
\begin{equation}
    r_\mathrm{sg} \simeq 2150 \alpha^{2/9}\left(\frac{M_\bullet}{10^9 M_\odot}\right)^{-2/9} \dot{m}^{4/9}\, .
    \label{eq:rsg}
\end{equation}
The viscosity parameter $\alpha$ was set to $0.1$ (in agreement with numerical studies, e.g. \citealp{Hawley2002,Hirose2009}) and 
$\dot{m}\equiv (\varepsilon_\mathrm{r}(0) / \varepsilon_\mathrm{r}(a)) (L/L_\mathrm{Edd})$.

The normalisation constants $\mathcal{A}$ and $\mathcal{B}$ were set so that (i) the total integrated luminosity equates to the AGN bolometric luminosity $L = \varepsilon_\mathrm{r}(a) \dot{M}_\bullet c^2$ and (ii) the relative normalisation between the optical and X-ray luminosities, which is characterised by the parameter $\alpha_{\rm OX}$,
\begin{equation}
    \alpha_{\rm OX} = 0.384 \log_{10}\left(\frac{L_\nu(\SI{2}{\kilo\electronvolt}/h)}{L_\nu(c/\SI{2500}{\angstrom})}\right)\, ,
\end{equation}
fits the physical models by \citet{Done2012} and \citet{Dong2012}, which roughly predict
\begin{equation}
    \alpha_\mathrm{OX}=-0.13\log_{10}(M_\bullet/M_\odot)+0.15\log_{10}(\fedd)-0.45.
\end{equation}
In practice, the spectrum at $\SI{2500}{\angstrom}$ and $\SI{2}{\kilo\electronvolt}$ is dominated respectively by the disc and the corona, which implies that the ratio between the normalisation constants is approximately given by $\mathcal{A}/\mathcal{B} \approx 10^{\alpha_\mathrm{OX}/0.384}(\SI{2500}{\angstrom}/\SI{2}{\kilo\electronvolt})^{\alpha_\mathrm{X}}$. Finally, the values of $\mathcal{A}$ and $\mathcal{B}$ can be obtained by integrating eq.~\ref{eq:SED_model} over frequency and imposing condition (i) above. We did not consider a reduction in the flux from the disc due to the random viewing angle. However, the correction would be on average of order unity and would only affect the UV fluxes.

In summary, this model depends on three BH parameters, $M_\bullet$, $f_\mathrm{Edd}$ and $a$, with only a weak dependence on the spin. A simplified version of this model was used in \citet{Trebitsch2021} to calculate the ionising radiation in \textsc{Obelisk}, therefore ensuring consistency between the in-simulation AGN properties and those calculated in post-processing. 
This SED is generally appropriate for radiatively efficient discs, which characterise BHs with $f_\mathrm{Edd} \gtrsim f_\mathrm{Edd,crit} = 0.01$. Otherwise, it can be regarded as an upper limit. We have checked that in our sample all BHs detectable in UV or X-ray and more than $94\%$ of those detectable in radio have $f_\mathrm{Edd}> f_\mathrm{Edd,crit}$ (see Sections~\ref{subsec:UV_X-rays} and~\ref{subsec:radio}).

We used this model to calculate the flux density in the rest-frame UV at $\lambda = 1500\,\si{\angstrom}$, which corresponds to optical to near-IR observed emission taking the redshift of sources, $z>3.5$, into account.  This is therefore the observability with optical telescopes.
We calculated the integrated flux in the observer-frame `total' X-ray band $[0.5-10]\,\si{\kilo\electronvolt}$. We further define the observer-frame soft, $[0.5-2]\,\si{\kilo\electronvolt}$, and hard, $[2-10]\,\si{\kilo\electronvolt}$, X-ray bands.

From a differential luminosity $L_\nu$, the spectral flux density at an observed frequency $\nu$ can be calculated from luminosity at the rest-frame frequency $\nu_\mathrm{em} = (1+z) \nu$,
\begin{equation}
    S_\nu = (1+z)\frac{L_{\nu_\mathrm{em}}}{4\pi D_{L}^2(z)}
\end{equation}
where the $(1+z)$ factor reflects the redshifting of the differential bandwidth ${\rm d}\nu$. The cosmological luminosity distance $D_L(z)$ of the source was calculated using the fiducial cosmology of the \textsc{Obelisk} simulation (see Section~\ref{subsec:Obelisk}).

The integrated flux in a given band can be calculated from the integrated rest-frame luminosity $L$ as follows,
\begin{equation}
    F = \frac{L}{4\pi D_L^2(z)}.
\end{equation}

\begin{figure}
    \includegraphics[width=\columnwidth]{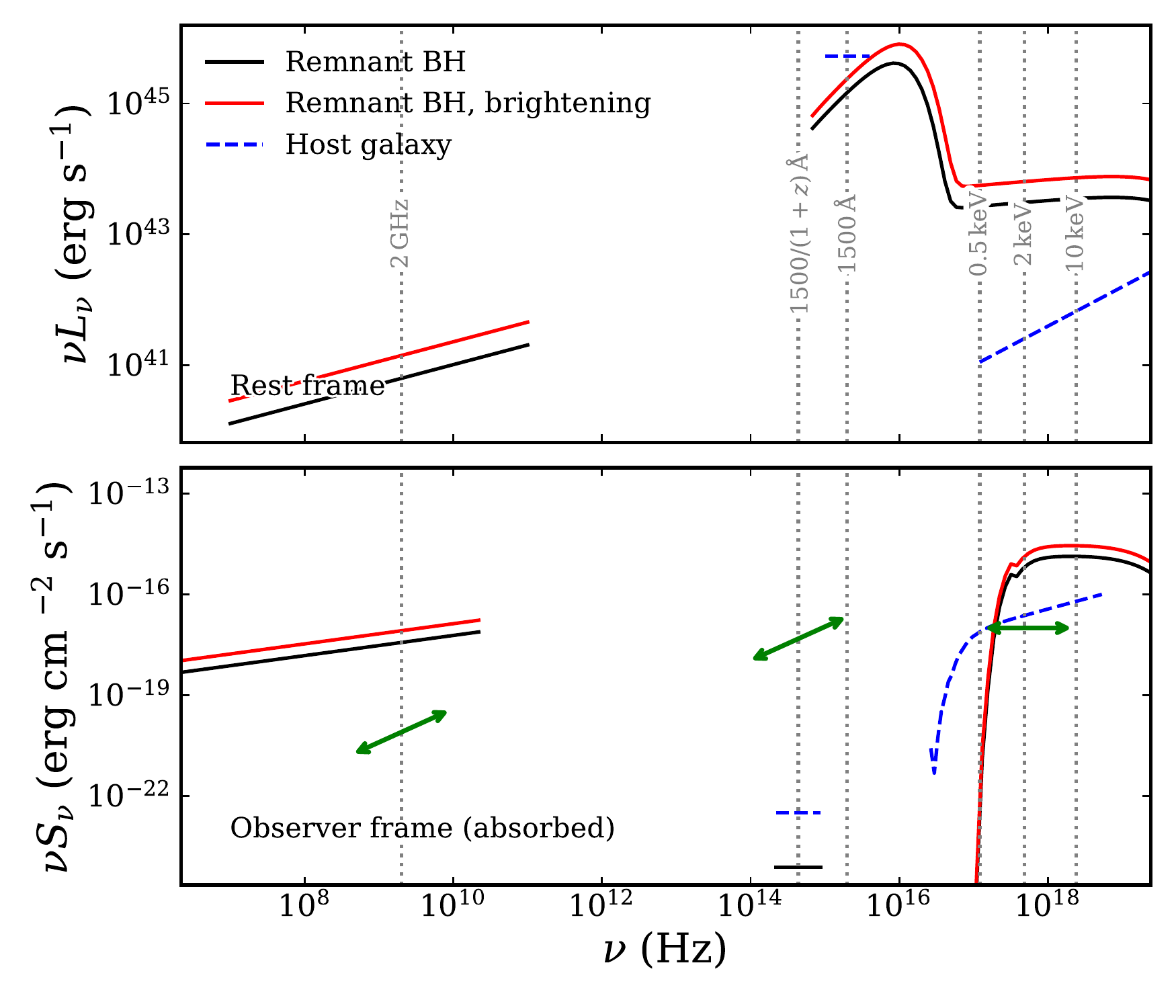}
    \caption{Spectral energy distribution of a merger remnant with mass $1.3\times10^8\,\Msun$, $f_\mathrm{Edd}=0.51$ and spin $0.90$, residing in a galaxy with mass $1.4\times10^{11}\,\Msun$ at redshift $3.55$. The top panel depicts the (unobscured) rest-frame luminosities while the bottom panel depicts the (obscured) observer-frame flux. The SED of the remnant BH is shown in the solid black and red lines for the fiducial and the brightening scenario, while the approximate UV and X-ray galaxy SED is shown in the blue dashed lines. The green arrows indicate approximately the instrumental sensitivities assumed in radio at $2\,\si{\giga\hertz}$ (assuming a sensitivity of $0.4\,\si{\micro\jansky}$), in UV at $1500\,\si{\angstrom}$ ($m_\mathrm{UV,lim}=31.3$), and in the $0.5-2\,\si{\kilo\electronvolt}$ band ($F_\mathrm{X,lim}=10^{-17}\,\si{\erg\per\second}$).}
    \label{fig:SED}
\end{figure}

We show an example of this SED model in Fig.~\ref{fig:SED} for a merger remnant with $M_\bullet=1.3\times10^8\,\Msun$, $f_\mathrm{Edd}=0.51$ and $a=0.90$, at $z=3.55$, and compare it with the emission from the host galaxy and some realistic instrumental sensitivity limits, which are defined in the sections below.

\subsection{AGN obscuration}
\label{subsec:AGN_obscuration}

We estimated the column density of gas in the interstellar medium (ISM) contributing to the BH obscuration by casting rays isotropically around each BH in the outputs of the simulation. For each BH we casted 100 rays and integrated the gas column density from the BH position to the virial radius of its host halo. We used a version of the \textsc{Rascas} code \citep{MichelDansac2020} modified to integrate column densities very efficiently.
For each sightline $i$, we computed $\exp(-N_{\mathrm{H},i} \sigma_{\rm T})$ and defined the typical column density around each BH as
\begin{equation}
\bar{N}_{\mathrm{H}} = -\ln\left(\left\langle \exp(-N_{\mathrm{H},i} \sigma_{\rm T}) \right\rangle\right).
\end{equation}

Additionally, we incorporated a subgrid model in order to account for the unresolved contribution from the $\mathrm{pc}$-scale, geometrically thick gas structure surrounding accreting BHs generally known as the torus. 
We assumed that the gas is at rest at infinity and that it falls radially under the action of the BH's gravitational pull. This assumption gives a lower limit on the time spent inside the torus by the accreted material, and therefore a lower limit on $N_\mathrm{H}$. We assumed a spherically symmetric configuration, which would roughly correspond to a spherically averaged problem. Under these assumptions, the gas density in the torus should follow a power-law density profile $\rho\propto r^{-3/2}$. We normalised the density so that the total mass in the torus, between its inner ($r_\mathrm{in}$) and outer ($r_\mathrm{out}$) radii is equal to $\dot{M}_\bullet t_\mathrm{ff}$. This is because the material crosses the torus in a free-fall time \citep[][]{Honig07} $t_\mathrm{ff} \sim \pi (r_\mathrm{out}^3/(8GM))^{1/2}$, where we have assumed $r_\mathrm{in} \ll r_\mathrm{out}$. The inner radius can be assumed to be the dust sublimation radius \citep{Suganuma06}, 
\begin{equation}
    r_\mathrm{in} \approx 0.47 \left(\frac{L_\mathrm{UV}}{10^{46}\,\rm erg\, s^{-1}}\right)^{1/2} \,  \si{\parsec}\, .
\end{equation}
The outer radius can be approximated as the BH gravitational sphere of influence for gas, that is, the BHL radius, at $r_\mathrm{out} = 2GM/(\bar{c_s}^2 + \bar{v}_\mathrm{rel}^2)$.

Further assuming that the infalling gas is only composed of hydrogen, we integrated the density in the radial direction to arrive at the following expression
\begin{equation}
    \bar{N}_\mathrm{H,torus} = \frac{3\pi}{4\sigma_\mathrm{T}} \frac{f_\mathrm{Edd}}{\varepsilon_{\rm r}} \frac{\bar{c_s}^2 + \bar{v}_\mathrm{rel}^2}{c} \left(\sqrt{\frac{r_\mathrm{out}}{r_\mathrm{in}}}-1\right).
\end{equation}
We note that the equation depends mainly on the Eddington ratio $f_\mathrm{Edd}$, but depends also on the BH mass, the radiative efficiency $\varepsilon_{\rm r}$, and the local gas properties.

In the X-rays, the rest-frame attenuated luminosity can be calculated as
\begin{equation}
    L_\mathrm{X,abs} = \int^{\nu_\mathrm{max}}_{\nu_\mathrm{min}} L_\nu e^{-\sigma_\mathrm{X}(\nu) N_\mathrm{H}} \, {\rm d}\nu \, .
    \label{eq:L_Xabs}
\end{equation}
The X-ray cross sections are calculated from the polynomial fits in \citet{Morrison1983}, extrapolated if needed to $\nu>\SI{10}{\kilo\electronvolt\per\h}$ assuming a scaling $\sigma_\mathrm{X} \propto \nu^{-3}$.

Given that \textsc{Obelisk} includes a model for dust, we followed a similar approach to estimate the UV obscuration. For each BH, we casted 100 rays in different directions. Along each sightline $i$, we computed the dust optical depth $\tau_{\mathrm{UV},i} = \kappa_{1500} \int_l \rho_{d,i} dl$ where $\rho_{d,i}$ is the local mass density of dust in each cell along the sightline $i$ and $\kappa_{1500}$ is the dust mass absorption coefficient at $1500\,\AA$. We estimated $\kappa_{1500}$ as follows: we started by assuming that our dust is composed of a mixture of silicate and carbonaceous grains with respective mass fractions 54\% and 46\% as in \citet{2018MNRAS.478.4905A}, following \citet{2009MNRAS.394.1061H}, and that the grain size distribution follows the MRN grain size distribution \citep{1977ApJ...217..425M} between $0.001$ and $0.25\,\mu\mbox{m}$. We then integrated the extinction cross sections from \citet{1993ApJ...402..441L} over the grain size distribution to get $\kappa_{1500}$. The typical attenuation is then defined as previously by
\begin{equation}
\bar{\tau}_\mathrm{UV} = -\ln\left(\left\langle \exp(-\tau_{\mathrm{UV},i}) \right\rangle\right).
\end{equation}

We similarly added a torus correction to the UV obscuration. In order to obtain a torus UV optical depth, we multiplied $N_\mathrm{H,torus}$ by a factor $\kappa_{1500}M_{\mathrm{dust},50}/M_{\mathrm{gas},50}$, where $M_{\mathrm{dust},50}$ and $M_{\mathrm{gas},50}$ are the dust and gas mass inside $r_{50}$.

The median value and interquartile scatter of the total gas column density for our numerical merger population is $\log_{10}(N_\mathrm{H}/\si{\per\cm\squared})=23.3\pm0.2$. The interstellar contribution ($\log_{10}(N_\mathrm{H,ISM}/\si{\per\cm\squared})=23.3\pm0.3$) dominates, while the torus contribution ($\log_{10}(N_\mathrm{H,torus}/\si{\per\cm\squared})=22.2\pm0.4$) only accounts for less than $10\%$ of the total median value. 
For dust, the median UV optical depth is $\log_{10}\tau_\mathrm{UV}=0.97\pm0.28$. In this case, the interstellar contribution still dominates ($\log_{10}\tau_\mathrm{UV,ISM}=0.84\pm0.31$), but the torus contributes more significantly ($\log_{10}\tau_\mathrm{UV,torus}=0.11\pm0.46$), to $30\%$ of the median value. We do not consider the contribution from the intergalactic medium since it is subdominant for our sample \citep{Arcodia2018}.

\begin{figure}
    \includegraphics[width=\columnwidth]{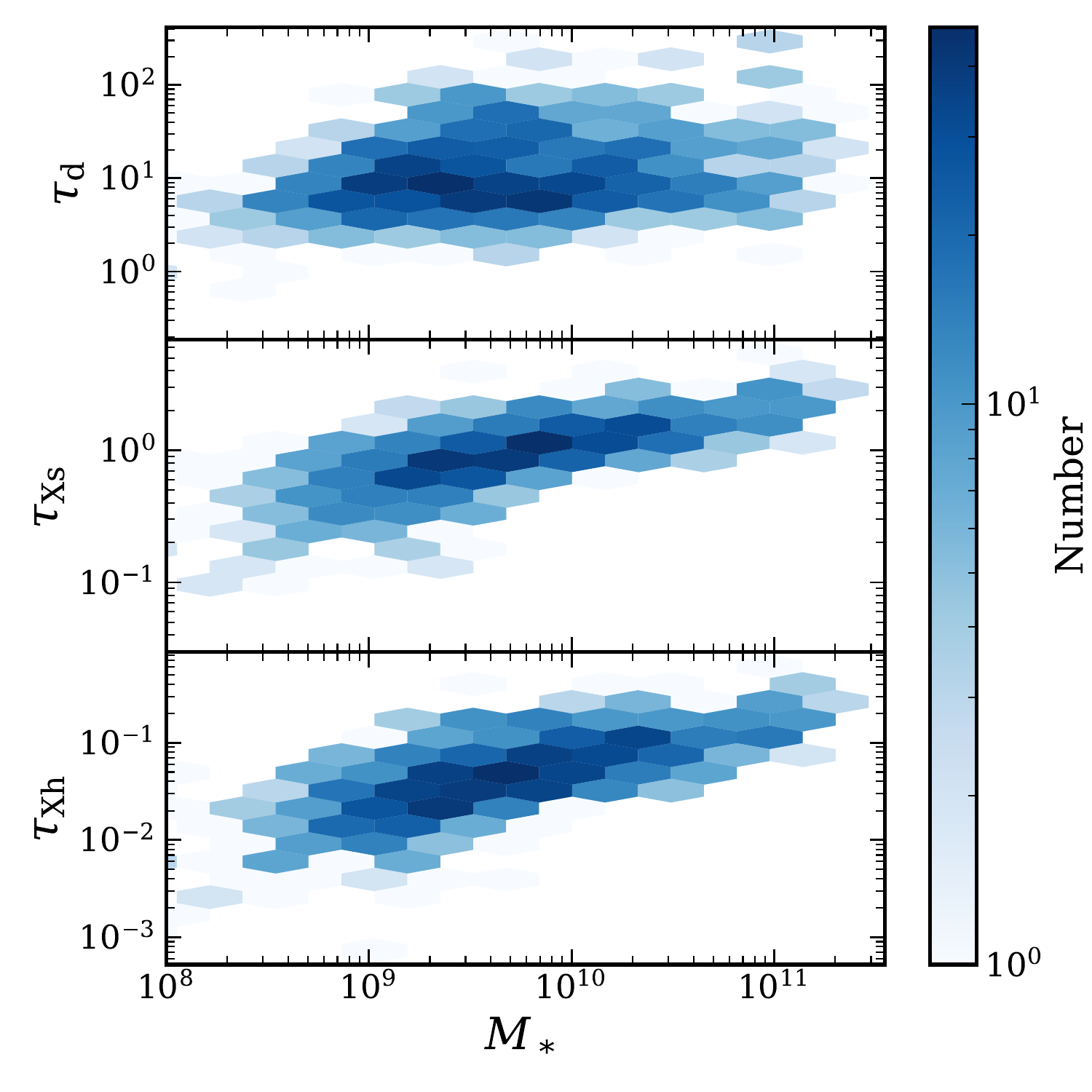}
    \caption{Distribution of the optical depth against the host galaxy stellar mass for remnant BHs in our numerical merger sample. Top panel: rest-frame UV; middle panel: observer-frame $0.5-2\,\si{\kilo\electronvolt}$; 
    bottom panel: observer-frame $2-10\,\si{\kilo\electronvolt}$. The colour denotes the number of BHs in each bin, on a logarithmic scale. Optical depths are very high in the UV, while in X-rays the optical depths are not extreme.}
    \label{fig:obscuration}
\end{figure}
We can define the mean optical depths in the observer-frame soft and hard X-rays $\tau_\mathrm{Xs}$ and $\tau_\mathrm{Xh}$ as $\tau=\ln{(F/F_\mathrm{abs})}$, where $F$ and $F_\mathrm{abs}$ are the integrated unabsorbed and absorbed flux. The distributions of $\tau_\mathrm{UV}$, $\tau_\mathrm{Xs}$, and $\tau_\mathrm{Xh}$ as a function of host galaxy mass $M_\ast$ are shown in Fig.~\ref{fig:obscuration} for remnant BHs in our sample of numerical mergers. The optical depths are generally very high in the UV. In the X-rays, the obscuration is much smaller, since at the high rest-frame frequencies probed the gas cross-sections are very small. Recall the simulation is limited at high redshift, so all our mergers occur at $z\gtrsim3.5$. 

The effect of obscuration can be seen in the bottom panel can be seen in Fig.~\ref{fig:SED}, where we show an example of an obscured observer-frame SED. The effect of obscuration is particularly strong in the UV.

\subsection{Radio emission}
\label{subsec:radio_method}

In order to assign a jet radio luminosity to the simulated mergers while bypassing the theoretical uncertainties related to the production of jets, we resorted to the `fundamental plane of black hole activity', an empirical correlation between the radio luminosity, X-ray luminosity, and mass of BHs. This relation has been shown to be applicable to BHs spanning 8 orders of magnitude in mass \citep[e.g.][]{Merloni2003,Falcke2004}. 
\citet{Gultekin2014} proposed that the fundamental plane relation found in \citet{Gultekin2009} also holds for a sample of low-mass ($M_\bullet \lesssim 10^{6.3}\,\rm \Msun$), highly accreting AGNs. More recent work \citep{Gultekin2022} has however shown that the fundamental plane tends to underestimate $L_\mathrm{R}$ at fixed $L_\mathrm{X}$ for a sample of highly accreting AGN powered by low-mass BHs. The fundamental plane only takes into account the contribution of the core radio luminosity. Therefore, we computed radio luminosity from the relation found in \citet{Gultekin2009}, but we treat this `pessimistic' model as a lower limit in the following analysis. We calculated the radio luminosity as
\begin{equation}
    \log_{10} L_{\mathrm{R},5\si{\giga\hertz}} = 4.80 + 0.78 \log_{10} M_\bullet + 0.67 \log_{10} L_\mathrm{X},
    \label{eq:FP}
\end{equation}
where $L_{\mathrm{R},5\si{\giga\hertz}} = \nu_{5\si{\giga\hertz}} L_{5\si{\giga\hertz}}$ is the radio luminosity at $\SI{5}{\giga\hertz}$, and $L_\mathrm{X}$ is the integrated X-ray luminosity in the rest-frame $\SI{2}{\kilo\electronvolt} - \SI{10}{\kilo\electronvolt}$ energy range. We calculated the radio luminosity at $\SI{2}{\giga\hertz}$ assuming a power-law spectrum $L_\nu \propto \nu^{\alpha_R}$ with index $\alpha_R=-0.7$ \citep{Gultekin2014}.

As an upper limit to the radio luminosity, we considered a theoretical model in which the jet is powered by the Blandford-Znajek effect \citep{Blandford1977}. We modelled the total synchrotron luminosity based on \citet{Meier2001},
\begin{equation}
L_\mathrm{S,s}=
    \begin{cases}
        \begin{aligned}
        &10^{46.0}\eta_\mathrm{S}\left(\frac{\alpha^\mathrm{AD}}{0.3}\right)^{-1}\left(\frac{M_\bullet}{10^9\, M_\odot}\right)\times \\
        & \left(\frac{f_\mathrm{Edd}}{0.1}\right) g^2 (0.55f^2+1.5fa+a^2) \,\si{\erg\per\second}
        \end{aligned} & f_\mathrm{Edd}<0.01\\
        
        \begin{aligned}
        &10^{43.5}\eta_\mathrm{S}\left(\frac{\alpha}{0.01}\right)^{-0.1}\left(\frac{M_\bullet}{10^9\, M_\odot}\right)^{0.9} \times \\
        & \left(\frac{f_\mathrm{Edd}}{0.1}\right)^{1.2} (1+1.1a+0.29a^2) \,\si{\erg\per\second}
        \end{aligned} & f_\mathrm{Edd}\geq0.01.
    \end{cases}
    \label{eq:jet_Meier}
\end{equation}
Here, the top equation represents the jet production from a geometrically thick advection-dominated accretion flow (ADAF) at low $f_\mathrm{Edd}$, while the bottom equation corresponds to the thin disc case. As above, we assumed for the thin disc $\alpha=0.1$, while for the ADAF we assumed $\alpha=0.3$ following \citet{Meier2001}. We set the dimensionless constant $f$ and $g$ to $1$ and $2.3$, following \cite{Meier2001} and \cite{Tamanini2016}. We further assumed, following \citet{Meier2001} that only a fraction $\eta_\mathrm{S}=10^{-2}$ of this power is radiated in the synchrotron spectrum. In contrast to the fundamental plane, this expression provides the total jet power, not only the core jet power. 

The radio luminosity $\nu L_\nu$ at $\nu=\SI{2}{\giga\hertz}$ can be calculated assuming the synchrotron radiation is emitted in a power-law spectrum with index $\alpha_{\rm R}$ over the frequency range $0.01-100\,\si{\giga\hertz}$. Overall, a fraction $\sim 10^{-3}$ of the initial power is transformed into radio luminosity $\nu L_\nu$ at $\SI{2}{\giga\hertz}$. In general, eq.~\ref{eq:jet_Meier} 
can predict radio luminosities more than $2$ orders of magnitude above the fundamental plane. Thus, we regard this `optimistic' model as an upper limit in the following analysis. 

We show an example of the optimistic radio model in Fig.~\ref{fig:SED}.

\subsection{Merger-induced brightenings and transients}
\label{subsec:transients}

A BH merger can potentially induce a brightening on a scale of days to years around the time of merger, which increases the luminosity of the remnant BH and constitutes a transient signal that can be used to detect the merger and identify it as such. 
To model this, we assumed that initially the remnant BHs emits in X-ray and UV at the fiducial luminosity predicted by our SED model above, for the appropriate accretion rate calculated in the simulation. That is, we assumed that the SED model for a single BH in an $\alpha$ disc applies. A brightening occurs either shortly before the merger \citep{Armitage2002,2016MNRAS.457..939C} or after $\sim t_\mathrm{vis}$, when the accretion of the inner rim drives a burst. We remain agnostic on the exact process, and we modelled a brightening by assuming $f_\mathrm{Edd}=1$ in our SED model\footnote{Assuming $f_\mathrm{Edd}=1$ in our SED model, which assumes a steady state solution, means assuming that the disc is more massive in general, at all radii. Having $f_\mathrm{Edd}=1$ accretion due to the `lump', means a large amount of material is crossing the horizon only at that particular time. That is, the amount of material is only large in the inner few gravitational radii where the material has piled up before the merger. On the other hand, the merger burst could also be super-Eddington \citep{Armitage2002}, therefore we consider our model as a reasonable `middle ground'.}. Physical arguments suggest that the formation of the cavity and the subsequent brightening should only happen if $q$ is large enough and if the binary is embedded in a gas-rich environment, but we optimistically assume this to apply to all mergers. On the other hand, we did not include a pre-merger suppression of the accretion rate, which is a pessimistic approach, in the sense that the luminosity change is weaker than if accretion were suppressed.

Different changes to jet properties around the time of BH mergers have also been proposed. Here, we considered the possibility of an increase in the jet radio luminosity due to an increase in the accretion rate analogous to the UV and X-ray model above. Again, we modelled this brightening by assuming $f_\mathrm{Edd}=1$ in our radio models.

We also considered a short-lived flare as a transient feature and modelled it as having, for an equal mass merger, a luminosity a factor $(1+\mathcal{K})$ higher than the pre-merger luminosity $L_\mathrm{S,s1}+L_\mathrm{S,s2}$. The increase in luminosity due to a flare, parametrised here by $\mathcal{K}$, is not well constrained by simulations \citep{Moesta2012,Gold2014b,Kelly2017,Cattorini2021,Cattorini2022}, and can depend strongly on the merger parameters. We assumed a factor of $\mathcal{K}=5$. Further, following the discussion in \citet{Kaplan2011}, we assumed an approximate scaling of the flare luminosity $L_\mathrm{S,f}\propto q^2M_\bullet^2$. For simplicity, we neglected the dependence on other parameters. In summary, we assumed that 
\begin{equation}
    L_\mathrm{S,f} = (1+\mathcal{K}q^2) (L_\mathrm{S,s1}+L_\mathrm{S,s2}).
    \label{eq:radio_flare}
\end{equation}

These two merger-induced transient phenomena (brightening and flare) will alter the small-scale core radio luminosity, but not necesarily the total radio luminosity. Therefore, we apply this radio transient model only on the `pessimistic model', which estimates only the core radio emission.

In our model, we did not consider the possibility explored by \citet{Yuan2021} and \citet{Ravi2018} that a gamma-ray burst-like source is produced as a newly formed jet after the merger impacts and shocks the nearby gas. This choice is based on simulations showing that the jet can exist both before and after the merger. Even in the case of a spin flip, \citet{Kelly2021} find that the jet direction does not change, remaining aligned with the ambient magnetic field on large enough scales. Similarly, \citet{Ruiz2023} do not find a significant perturbation in the jet propagation due to the change in spin direction for initially slowly spinning BHs. This presumably means that the BH will eventually realign with the jet, although the outcome is unclear and depends on complex physics \citep[see][]{McKinney2013,Liska2021}. In this case the jet will continue to propagate in the same direction as before, rather than encountering pristine gas. 

Finally, we note that other types of transient features can appear around the time of the merger: spectral changes caused by perturbations in the accretion disc \citep{2008ApJ...684..835S}, complex lightcurves in the case of kicked BHs \citep{2010MNRAS.401.2021R}, periodic modulations \citep{2022ApJ...928..137G}. Exploration of these and also of features occurring during the inspiral \citep{2012MNRAS.420..860S,2014ApJ...785..115R,Farris2015} are postponed to a future investigation.

\subsection{Galactic emission}
\label{subsec:galactic_emission}

We also modelled the galactic emission, which in our case acts as contamination hindering the detection of the BH merger.
In the UV band, the galactic emission was computed from the stellar population in the galaxy. For each galaxy in our catalogue, we estimated the intrinsic UV luminosity at $1500\,\mbox{\AA}$ from the properties of the star particles in the galaxy. Each star particle was attributed a luminosity based on its age and metallicity using the \textsc{Bpass} v2.2.1 SED \citep{Eldridge2017,Stanway2018} and rescaled to the mass of the star particle. The intrinsic luminosity of the galaxy was then obtained by summing over all star particles associated with the galaxy. 
As \textsc{Obelisk} includes a model for the formation, growth, and evolution of dust in the galaxy, we can estimate the observed UV luminosity by computing the average attenuation. For this, we used the \textsc{Rascas} code to cast 100 rays isotropically from each star particle within $5 r_{50}$ of each galaxy. We measured the dust optical depth along each ray and used the average $f_{\rm UV} = \langle \exp(-\tau_{\mathrm{UV},i})\rangle$ as the escape fraction of UV light. The observed UV luminosity of a galaxy was then defined as the intrinsic luminosity times $f_{\rm UV}$. We note that our method does not account for orientation effects.

The galactic X-ray emission was assumed to be dominated by X-ray binaries (XRBs). We modelled the X-ray luminosity of XRBs using the empirical scaling relation found in \citet{Fornasini2018}, in which the integrated X-ray luminosity in the $\SI{2}{\kilo\electronvolt}$ -- $\SI{10}{\kilo\electronvolt}$ range is parametrised as a function of the galaxy stellar mass $M_\ast$ and star formation rate ($\mathrm{SFR}$) as follows,
\begin{multline}
    L_\mathrm{XRB} = 10^{29.98}(1+z)^{0.62}\left(\frac{M_\ast} {M_\odot}\right)  \,\si{\erg\per\s}+\\
     10^{39.78}(1+z)^{0.2}\left(\frac{\mathrm{SFR}}{M_\odot\,\rm yr^{-1}}\right)^{0.84} \,\si{\erg\per\s}.
     \label{eq:L_XRB}
\end{multline}
The effect of gas absorption was added assuming a constant column density of $N_{\rm H}=10^{22} \,\rm cm^{-2}$ and a power-law spectrum with photon index $\Gamma = 1.4$. 

We compare the host galactic emission with the remnant BH of a particular merger in Fig.~\ref{fig:SED}. In general, the galactic emission can be comparable with the AGN emission.

The radio emission generated by star-forming regions can also hinder the radio detectability of AGN. As an order of magnitude estimate, we calculated the galactic radio emission from the scaling relations in \cite{2003ApJ...586..794B}, which relate the radio luminosity $L_{\mathrm{R,gal}}$ to the $\mathrm{SFR}$. Since those estimates are based on a fit of $\mathrm{SFR}$ as a function of $L_{\mathrm{R,gal}}$ and not the converse, which is what we need, and the data is limited to low redshifts, we do not include them explicitly in our analysis, but we note that the SFR-induced radio luminosities could be comparable or higher than the AGN for BHs with mass $M_\bullet\lesssim 10^{7.5}\,\Msun$. 

\subsection{LISA gravitational wave analysis}

We calculated the detectability and parameter estimation for the satellite LISA for our set of simulated mergers. Since delayed mergers do not have a well-defined mass ratio, as it is unclear how much mass is accreted onto which black hole during the sub-grid inspiral, we restricted our GW analysis to the sample of numerical mergers.

To simulate the GW signal, we adopted the PhenomHM waveform \citep{PhHM} that assumes spins aligned with the binary orbital momentum but includes higher order harmonics to break degeneracies in the parameter estimation process.
The GW signal from a BH-BH binary with aligned spins can be described by 11 parameters: the primary and secondary mass $M_1$ and $M_2$ (with $M_1>M_2$), the component of the spins aligned to the orbital angular momentum $\chi_1$ and $\chi_2$ ($\chi_{1,2}=\vec{a_{1,2}}.\vec{\ell}/\ell$), the time of coalescence $t_c$, the luminosity distance $D_L(z)$, the inclination $\iota$, the sky latitude $\beta$ and longitude $\lambda$, the orbital phase at coalescence $\phi$, and the polarisation $\psi$. We can additionally define the chirp mass $M_{\rm chirp} = (M_1M_2)^{0.6}(M_1+M_2)^{-0.2}$.
$M_1$, $M_2$, $\chi_1$ and $\chi_2$ are directly produced in the simulation. 
The sky latitude $\beta$ and longitude $\lambda$ were set randomly over the sphere as well as the inclination $\iota$ and the polarisation $\psi$. The phase at coalescence $\phi$ is randomised between $[0,2\pi]$. The time to coalescence $t_c$ was set randomly between $[0,1]$ yr.

For a single event, we computed the signal-to-noise ratio ($\mathrm{S/N}$) as
\begin{equation}
    \label{eq:S/R}
    {\rm S/N} = \int_{\nu_{\rm lo}}^{\nu_{\rm up}} \frac{|\tilde{h}(\nu)|^2}{S_n(\nu)}{\rm d}\nu \, ,
\end{equation}
where $\tilde{h}(\nu)$ is the Fourier transform of the time-domain GW signal, $S_n(\nu)$ is the noise power spectral density, and $\nu_{\rm lo}$ and $\nu_{\rm up}$ are the minimum and maximum frequencies of integration. For $S_n(\nu)$ we adopted the so-called `SciRDv1' sensitivity \citep{Stas_sensitivity} and we set $\nu_{\rm lo} = 10^{-5} \, \rm Hz$ and $\nu_{\rm up} = 0.5 \rm \, Hz$. If $t_c$ is very short, the initial frequency $\nu_{\rm lo}$ was reduced. We also added to the LISA power spectral density the noise from the population of unresolved galactic binaries \citep{2021PhRvD.104d3019K}, with an amplitude corresponding to three years of observations.
The posterior distributions on the binary parameters $\bar{\theta}$ can be obtained following Bayes theorem as
\begin{equation}
    p(\bar{\theta}|d) = \frac{\mathcal{L}(d|\bar{\theta}) \pi(\bar{\theta})}{p(d)}\,, 
\end{equation}
where $\mathcal{L}(d|\bar{\theta})$ is the likelihood of observation $d$ with parameters $\theta$, $\pi(\bar{\theta})$ are the prior probabilities on the binary parameters and  $p(d) = \int {\rm d} \bar{\theta} \mathcal{L}(d|\bar{\theta}) \pi(\bar{\theta})$ is the evidence.

The binary posterior distributions were obtained following the formalism presented in \cite{Marsat21}. For each binary, we ran the Bayesian Markov Chain Monte-Carlo (MCMC) analysis for 2000 iterations with 64 walkers and 10 temperatures. 
We added an additional step to our analysis. Since we are interested in the sky localisation $\Delta \Omega$ to detect the possible EM emission, we decided to re-run the systems with $5<\Delta \Omega / {\rm deg^2}<40$ for $10^5$ iterations to ensure the convergence of the MCMC algorithm.

\begin{figure*}
    \includegraphics[width=\textwidth]{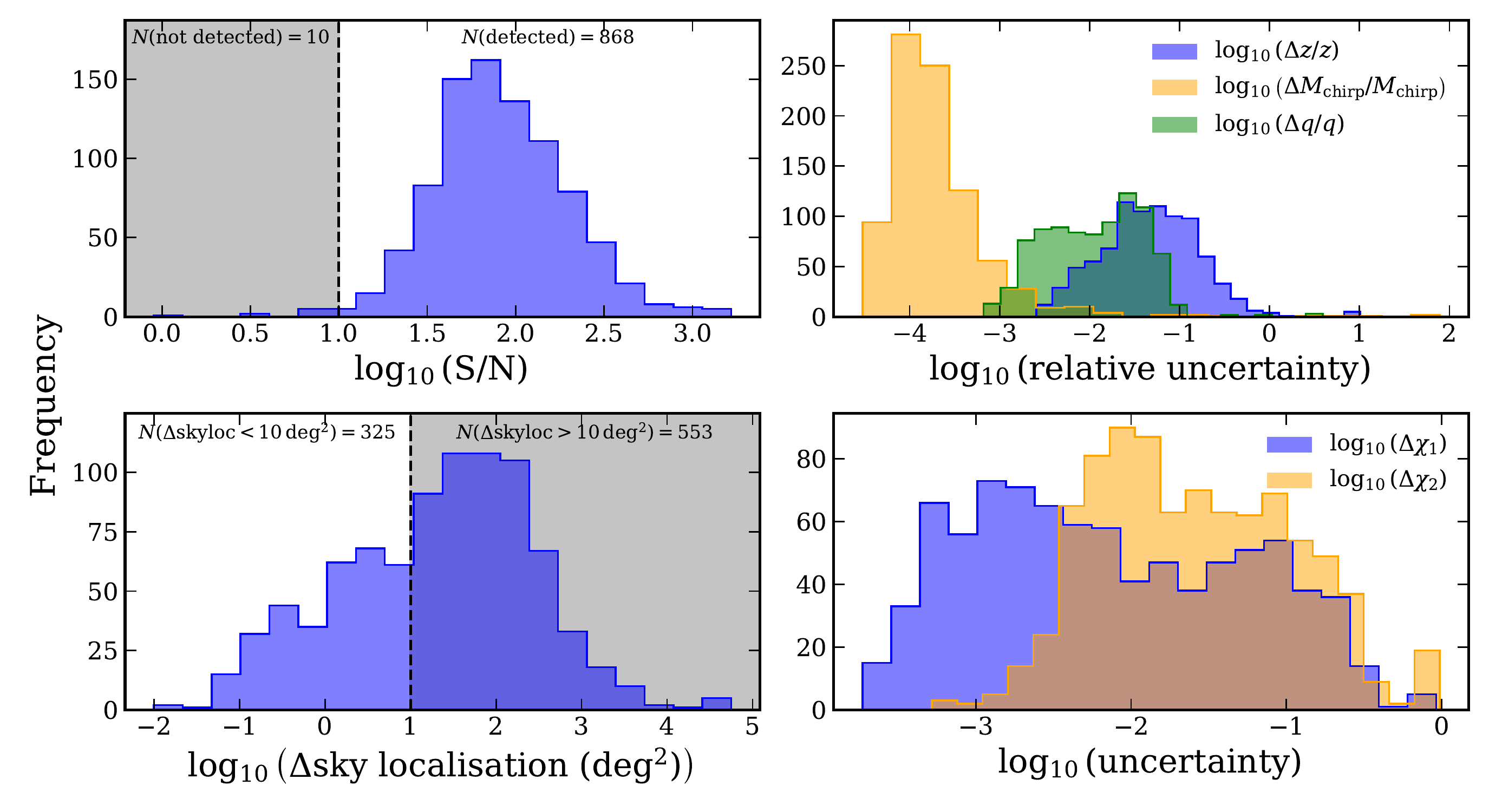}
    \caption{Parameter distribution for the mock LISA data analysis. The top-left panel shows the signal-to-noise ratio ($\mathrm{S/R}$) distributions. Gravitational wave events are assumed to be detected if $\mathrm{S/R}>10$, indicated by the vertical dashed line. The $90\%$-confidence error in the sky localisation is shown in the bottom-left panel. The dashed line delimits the shaded region, which corresponds to the events with sky localisation poorer than $10 \, \mathrm{deg}^2$. The relative $90\%$-confidence error distributions for $z$, $M_\mathrm{chirp}$ and $q$, and for $\chi_1$ and $\chi_2$, are shown in the right panels, top and bottom respectively. Most parameters are recovered with small uncertainties. The majority of mergers have sky localisation worse than $10\,\mathrm{deg}^2$ at the time of the merger.}
    \label{fig:GW_analysis}
\end{figure*}

\section{Multi-messenger observability of BH mergers}
\label{sec:observability}

In this section, we analyse the detectability of merging BHs at all redshifts probed by the simulation, $z \geq 3.5$. We analyse the GW emission of the sample of numerical mergers. We recall that we consider only numerical mergers because the mass ratio $q$, which is crucial for GW data analysis, is not well-defined for delayed mergers (see Section~\ref{subsec:merger_selection}). The analysis of EM signals is extended to both numerical and delayed mergers. 

We define a number of samples that are favourable for an EM detection. We define a source as \textbf{AGN-dominated} if the flux from the BH is larger than the flux from its host galaxy, and so the galactic contamination does not hinder the detection of the BH. We denote the converse case, where the galactic flux dominates the BH flux, as \textbf{galaxy-dominated}. A BH is considered \textbf{observable} if its flux exceeds the instrument sensitivity and the source is AGN-dominated. If the GW analysis is performed for the sample, we additionally require that a GW signal be detectable by LISA to consider it observable as a multi-messenger source. We note that this definition does not require an EM transient to be present.

We define an \textbf{EM counterpart} to a GW event as a source that exhibits a merger-induced transient with a significant change in flux, enabling identification as a merger. We assume that such a merger-induced variation in the flux can be detected if either: (i) The source `appears' at the time of the transient -- it is undetected before the transient and detected at the transient. The BH is observable at the transient. (ii) The source `disappears' -- it is detected before and undetected after. The BH is observable before. (iii) The source is detected before and after, but the flux changes significantly. In this case (iii), we consider a flux difference to be significant enough if the flux varies by more than a factor of $2$. This is probably an optimistic choice since AGN are intrinsically variable sources. Additionally, the BH must be observable either before or at the transient. We denote mergers that fall into any of these categories as having an EM counterpart. In the following, we do not consider transients of type (ii) since in the analysis below we do not find any transients of this type.

\subsection{GW observability and LISA parameter estimation}
\label{subsec:GW_observability}

We first discuss the detectability and parameter estimation by the LISA satellite. The analysis is performed by accumulating signal from the time the binary enters the LISA band up until coalescence.
Fig~\ref{fig:GW_analysis} shows for the numerical merger population the distribution of the $\mathrm{S/R}$, and the distribution of $90\%$-confidence uncertainties in estimating $z$, $M_\mathrm{chirp}$, $q$, $\chi_1$ and $\chi_2$, and the sky localisation. We find that most of the mergers in our sample are detectable if we set the limit for detectability at $\mathrm{S/R}>10$. This is because the masses of our BH population generally lie within the target range of LISA, and the mass ratios are generally moderate. An exception is the lowest mass ratio mergers ($q\lesssim10^{-3}$), which in our simulation correspond to mergers with a total mass of $M_\bullet\gtrsim10^8\,\si{\solarmass}$. These mergers have low $\mathrm{S/R}$ despite occurring at the lowest redshifts of our sample, $z<3.7$. Additionally, some high-redshift ($z\gtrsim4$) edge-on mergers of BHs close to the seed mass ($M_\mathrm{chirp}\lesssim 10^5\,\si{\solarmass}$) are undetected or detected with poor $\mathrm{S/R}$.

\begin{figure*}
	\includegraphics[width=\textwidth]{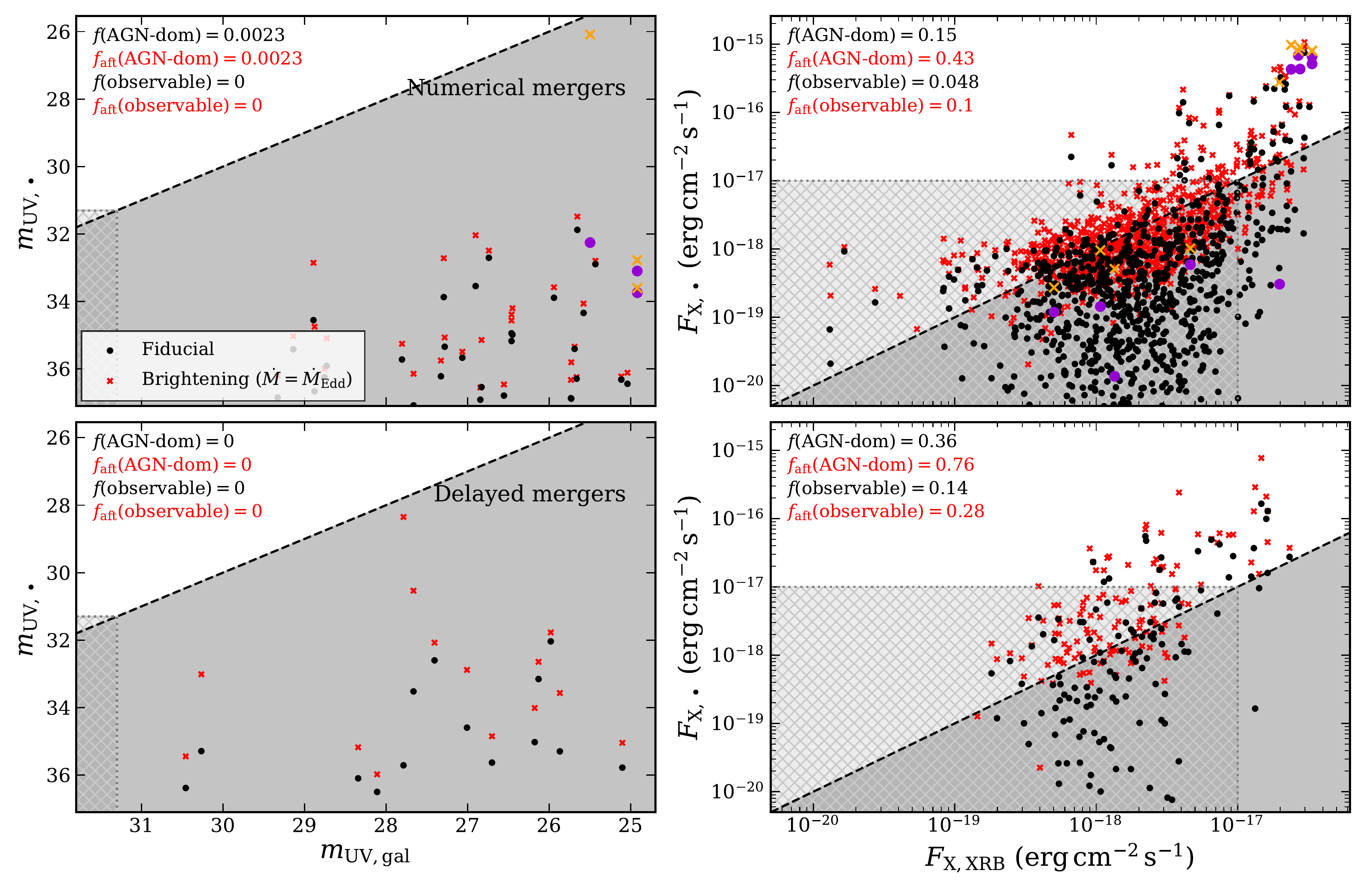}
    \caption{Black-hole vs galaxy flux, both corrected for absorption.  Left column: rest-frame UV magnitude at $\nu=\SI{1500}{\angstrom}$; right column:  X-ray $0.5-10\,\si{\kilo\electronvolt}$ band. Top row: numerical mergers; bottom row: delayed mergers. Black dots: accretion rate measured in the simulation; red crosses:  merger-induced brightening with $f_\mathrm{Edd}=1$. The grey region corresponds to galaxies dominating the emission over their BHs. The black dashed line indicates the boundary where the luminosity and galaxy and BH are equal.
    The hatched region corresponds to galaxies and BHs below the minimum flux needed for detection, which we set to $m_\mathrm{UV,lim}=31.3$ (UV) and $F_\mathrm{X,lim}=10^{-17}\,\si{\erg\per\centi\meter\squared\per\second}$ (X-rays). In the top left corner, we indicate the fraction of BHs with higher flux than their host galaxies and that of BHs above the observational flux limit. We note that a significant fraction of sources are very faint and lie beyond the limits of the plot, especially in the UV. Mergers that are GW undetected by LISA are shown with bigger and lighter markers, in purple and orange, for the fiducial and brightening cases. In UV all merging BHs are outshone by their host galaxy, while in X-rays a fraction of merging BHs are both brighter than the host galaxy and detectable by sensitive X-ray telescopes.
    }

    \label{fig:L_BH_vs_L_gal}
\end{figure*}

The parameters encoding the binary masses, $M_\mathrm{chirp}$ and $q$ are estimated with high precision, especially $M_\mathrm{chirp}$. The redshift and the spins are recovered with good precision, but the distributions have tails extending to high uncertainties that may become comparable to the value of the parameters.
The $90\%$-confidence uncertainties in the sky localisation at merger, which corresponds to the best available estimate \citep[compared to considering the inspiral phase only, see][]{2020PhRvD.102h4056M}, are generally poor -- only $37\%$ of the mergers have a localisation better than $10\,\mathrm{deg}^2$. The redshift determination has typical uncertainty of 0.01-0.1, which means that the 3D error box is dominated by the sky localisation uncertainty. As a reference, in COSMOS2020 \citep{Weaver2022} at $i-$magnitude $< 27$ and $3.49<z<3.51$ there are $740$ galaxies per $\mathrm{deg}^2$ (M. Shuntov, private communication).

This hinders the possibility of using GW detection to guide the search for EM counterparts with instruments having a small field of view. In radio, the field of view of ngVLA and SKA are below $2\,\mathrm{deg}^2$ so in most cases several pointings are needed in order to cover the LISA sky localisation error-box. In X-ray the field of view of Athena is planned to be $0.4\,\mathrm{deg}^2$ ($0.25\,\mathrm{deg}^2$ for NewAthena) and that of AXIS has a proposed $\sim 0.13\,\mathrm{deg}^2$, which will require more tiling, while the NASA Transient Astrophysics Probe is proposed to be $1 \,\mathrm{deg}^2$, but with a lower sensitivity than Athena and AXIS. THESEUS has a very large field of view, 0.5-2 sr, but it is expected to have much lower sensitivity.  In optical, with large field-of-view instruments such as the Rubin Observatory, $9.6\,\mathrm{deg}^2$ \citep{Ivecic2019}, one can use only a few tiles to cover the error-box.

The uncertainty in the sky localisation of our systems is mainly determined by the inclination $\imath$, the angle of the orbital angular momentum of the binary with respect to the line of sight. Face-on systems are better localised, leading to an error of $\sim 10^{-1}$ on average, while edge-on systems lead to $\sim 10^3$. Since our systems are distributed uniformly in orientation, the inclination angle $\imath$ is randomly distributed with a probability proportional to $\sin \imath$. This results in a large scatter in the sky localisation with values preferentially skewed towards the poorly localised regime. It is important to note that the waveforms used in our parameter estimation do not include the effects of spin precession. The inclusion of spin precession could improve the errors by a factor of $2-5$ \citep{2006PhRvD..74l2001L}, which in our case would lead to $46\%-60\%$ of the mergers being localised better than $10\,\mathrm{deg}^2$ if all errors were scaled down equally. Our current value of $37\%$ is likely a lower limit.

If merger parameters are estimated from the median of our parameter distributions, we generally recover the true values with good accuracy. However, it is worth noting that there is a small number of extreme outliers in our sample. These are generally events detected with low $\mathrm{S/R}$, for which the parameter estimation returns large errors and uncertainties. These outliers tend to strongly overestimate the luminosity distance (reaching values of up to $z\sim40$ assuming the fiducial cosmology) and chirp mass of the events, which may be interpreted erroneously as evidence for massive primordial black holes \citep[see][for detailed models of how GWs can constrain primordial black holes]{2021JCAP...11..039D,2022ApJ...931L..12N,2022JCAP...08..006M}. In all these cases, the high deviations are accompanied by higher uncertainties, which offer a way to flag them as outliers.

\begin{figure}
    \includegraphics[width=\columnwidth]{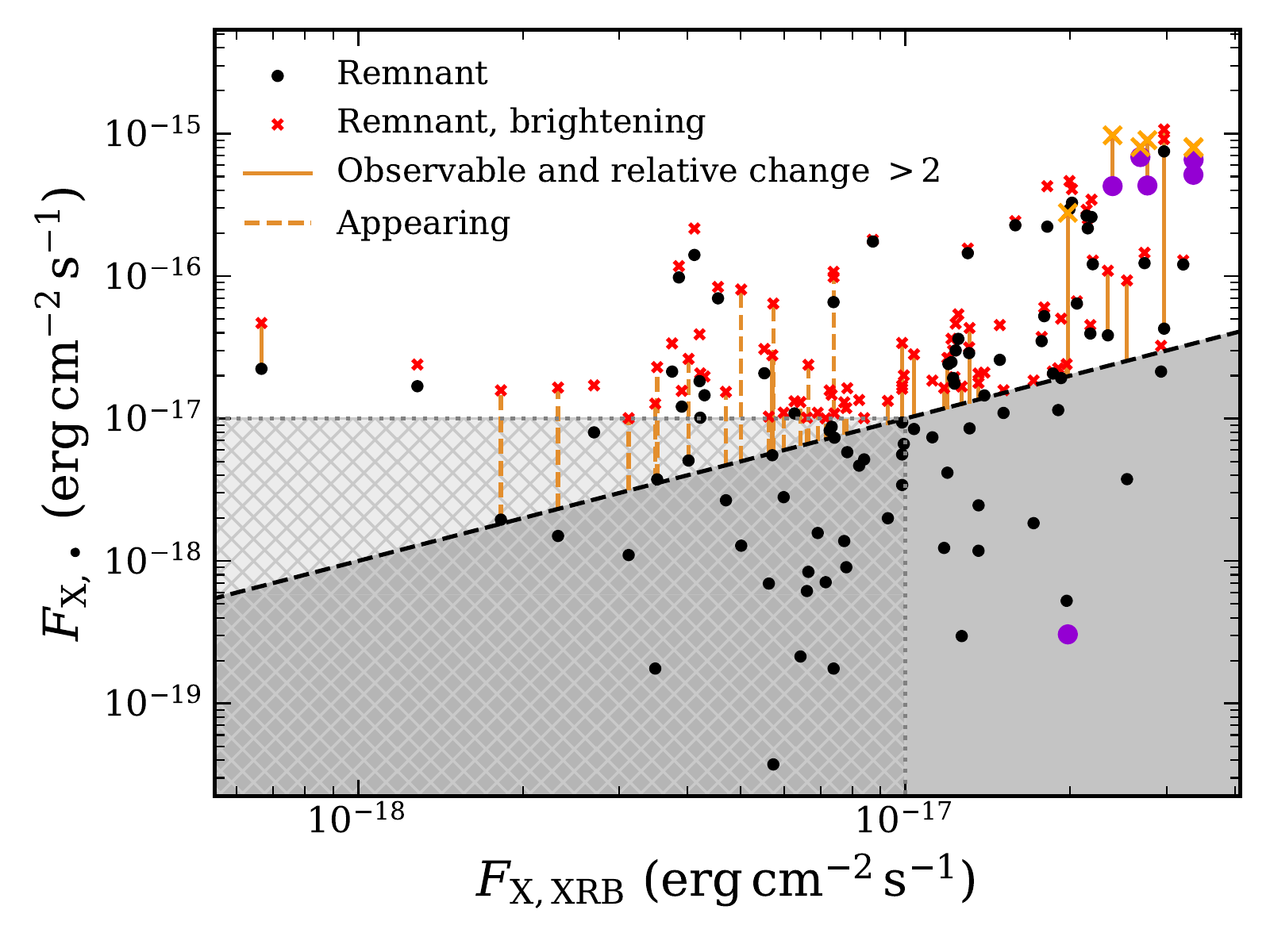}
    \caption{X-ray transients in the numerical merger sample. Similar to the right panel of Fig.~\ref{fig:L_BH_vs_L_gal}. Black dots and red crosses correspond to the X-ray post-merger BH fluxes for the pre-transient case (fiducial properties) and the brightening (which assumes $f_\mathrm{Edd}=1$). We only show mergers for which either of the two configurations is observable. Pre-transient and brightening fluxes are connected by orange dashed lines if the pre-transient source (total flux from the BH and the galaxy) is undetected and the brightening source is detected and the BH is observable, and orange solid lines if both fluxes are detectable and the flux change is larger than a factor of $2$. GW-undetectable remnant BHs are shown in purple (pre-transient) and orange (brightening).}    
    \label{fig:Xray_transients}
\end{figure}

\begin{figure}
    \includegraphics[width=\columnwidth]{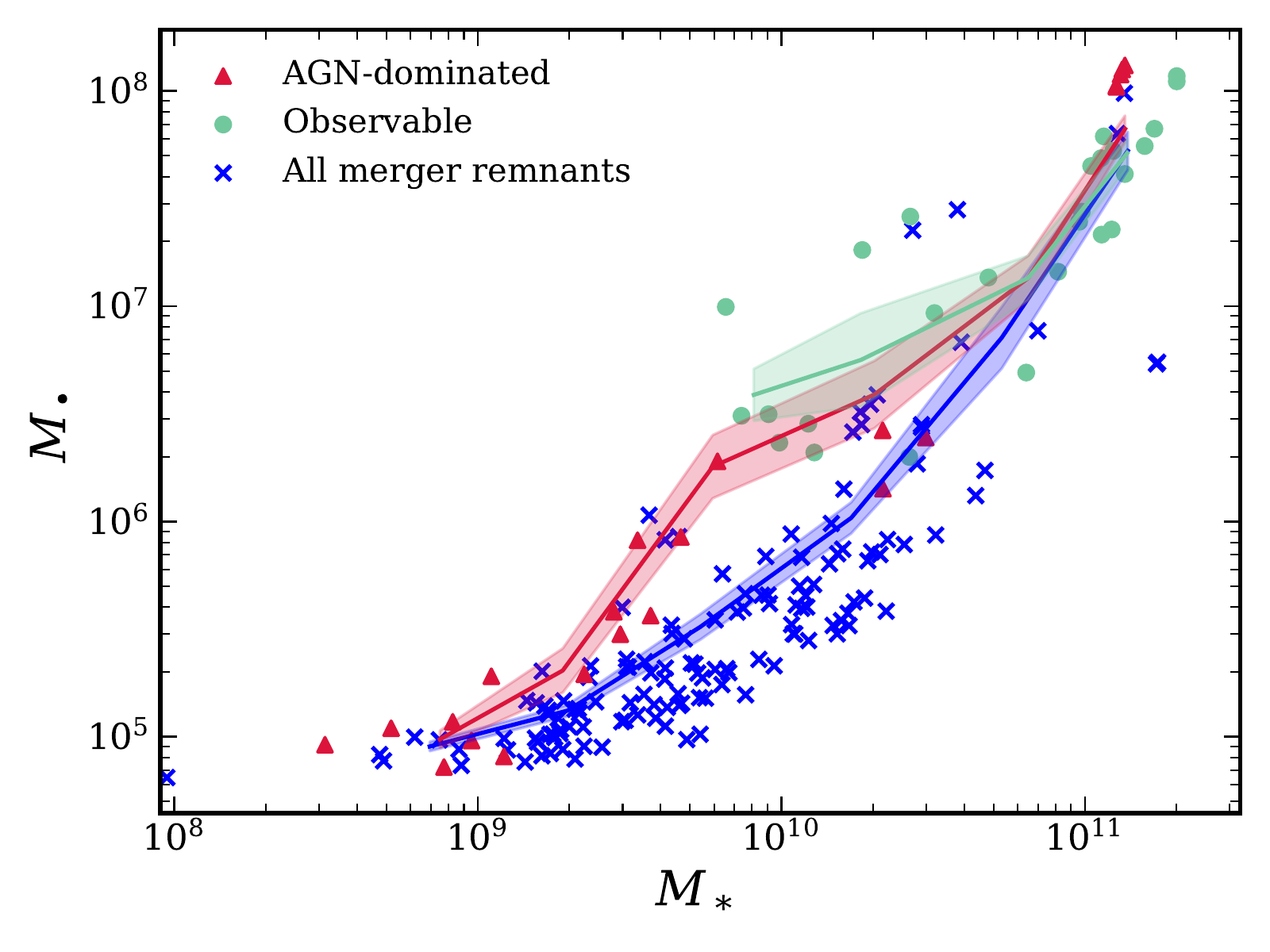}
    \caption{Mass of the remnant BH against the mass of the host galaxy for mergers occurring in the range $z=3.5-4$. Blue crosses correspond to the overall population of numerical remnant BHs, crimson triangles to AGN-dominated mergers and green circles to observable mergers in the X-ray. Lines denote the geometric average of $M_\bullet$ in bins of $M_\ast$, while shaded regions denote the $1\sigma$ error in the average. We note that all observable mergers are AGN-dominated sources. AGN-dominated mergers are more massive, at fixed $M_\ast$ compared to all merger remnants: this is an effect of requiring the AGN to be bright, and brighter than the galaxy. This effect is stronger for observable mergers. A similar trend is found at all redshifts.}
    \label{fig:MBH_vs_Mgal_Xray}
\end{figure}

\subsection{UV and X-rays}
\label{subsec:UV_X-rays}

\subsubsection{UV and X-ray detectability}
\label{subsubsec:UV_X-rays_detectability}

In the following sections, we explore the possibility of an EM detection that would complement the GW detections discussed in the previous section.
In Fig.~\ref{fig:L_BH_vs_L_gal}, we show for numerical and delayed merger remnants the attenuated remnant AGN flux in the rest-frame UV ($1500\,\AA$) and the observer-frame $0.5-10\,\si{\kilo\electronvolt}$ band X-rays against the host galaxy flux, which in our case acts as a contaminant hindering the detection of the central AGN. The rest-frame UV wavelength considered corresponds to the optical or near-infrared in the observer frame for our high-redshift sample -- at $z=3.5$ it would be observed in the $g$-band, while at $z=7$ it would be observed in the $z$-band. For the modelling of the AGN SED, the implementation of obscuration, and the estimation of the galactic emission, we refer to sections~\ref{subsec:SED},~\ref{subsec:AGN_obscuration},~and~\ref{subsec:galactic_emission}, respectively. For our merger hosts, the XRB luminosity is generally dominated by high-mass XRBs, whose luminosity depends on the $\mathrm{SFR}$.  

\begin{figure*}
    \includegraphics[width=\textwidth]{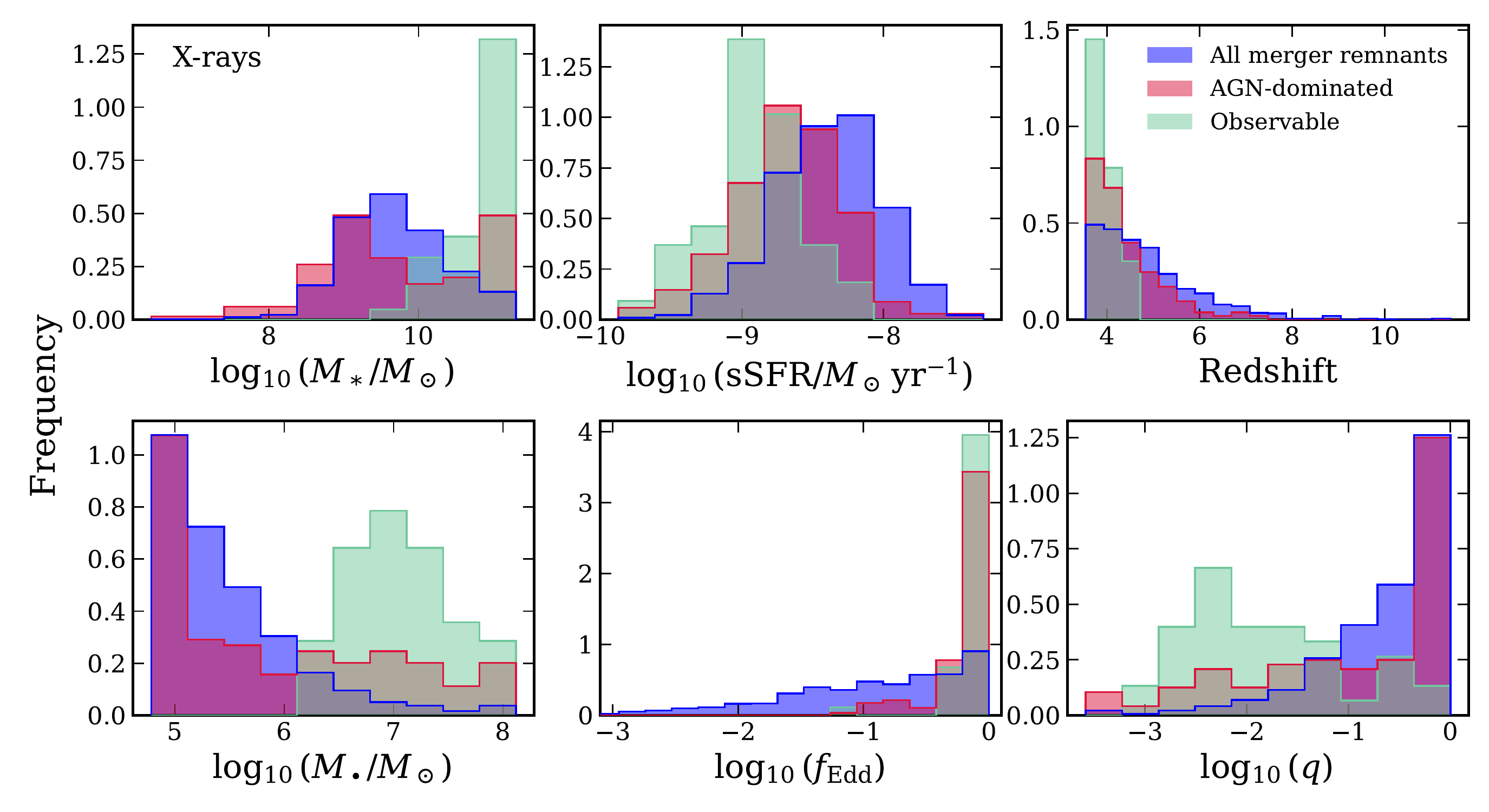}
    \caption{Distribution of $M_\ast$, $M_\bullet$, $\mathrm{sSFR}$, $f_\mathrm{Edd}$, redshift and $q$ for the overall population of numerical BH mergers (in blue) and for the sub-sample of AGN-dominated mergers (in crimson), and observable mergers (in green) in the X-rays. The integrals of all distribution functions are normalised to unity. X-ray-observable merger remnants accrete faster and are more massive with respect to the general merger remnant population. They occur at lower redshifts and are hosted in more massive galaxies with lower $\mathrm{sSFR}$.}
    \label{fig:Xray_pops}
\end{figure*}

Recall that we refer to sources in which the remnant BH dominates its host as AGN-dominated, while we refer to the converse case where the host dominates over the BH emission as galaxy-dominated. We also set optimistic observability thresholds, at $m_\mathrm{UV,lim}=31.3$ in the UV \citep[corresponding to the sensitivity for isolated point sources to 5$\sigma$ in 5 hours of the MICADO instrument on ELT\footnote{The field of view of MICADO is $\sim 50\times50\,\mathrm{arcsec}^2$, so the source requires prior identification with a different instrument.},][]{Davies2010}, and $F_\mathrm{X,lim}=10^{-17}\,\si{\erg\per\centi\meter\squared\per\second}$ in the X-rays, a reasonable flux limit for missions such as AXIS \citep{Mushotzky2018} and Athena \citep{Nandra2013}. We refer to BHs that exceed this limit and dominate the galactic emission as observable. For numerical mergers, we additionally require the merger to be detectable by LISA for it to be considered observable as a multi-messenger source. We note that our definition of observable does not require a transient to be present.
 
We find virtually no observable merger remnant BHs in the restframe UV. The few cases that exceed the observability threshold are galaxy-dominated because the high-redshift galaxies in our sample are actively star-forming. Even in the presence of a merger-driven brightening we find that it remains unlikely to detect our sample of merger remnants. If we do not consider dust obscuration, which is very high for AGN in the UV (see Fig.~\ref{fig:obscuration}), all sources also remain galaxy-dominated.

In X-rays, the detectability of BH mergers is more favourable. A fraction $\sim 15\%-35\%$ ($134$ out of $878$ -- $46$ out of $129$) of sources are AGN-dominated, depending on whether numerical or delayed mergers are considered. This is because the galactic emission is lower relative to the BH emission in X-rays compared to UV, and gas obscuration is only a small effect due to the small gas cross-sections at the very high energies probed ($0.5(1+z)-10(1+z)\,\si{\kilo\electronvolt}$ in the rest-frame X-ray, see Fig.~\ref{fig:obscuration}). A fraction $\sim 5\%-14\%$ ($42$ out of $878$ -- $18$ out of $129$) of all BH mergers is observable. Most BHs above the observability limit also dominate over the galactic emission. Unfortunately, some of the brightest mergers in the X-rays are not observable in GWs. These events correspond to a population of high-mass, low-$q$ mergers, which have poor detectability.

The fraction of observable remnant AGN in X-rays is larger for delayed mergers compared to numerical mergers. Delayed mergers remnants tend to be more massive compared to numerical mergers since the galaxy merger can induce an accretion burst that increases the mass of the BHs and delayed mergers are biased towards high-mass galaxies that can experience mergers early \citep{Dong-Paez2023a}. A more massive BH population leads to a higher number of observable mergers. The fraction of AGN-dominated is also higher for delayed mergers, for two reasons: firstly, delayed mergers tend to have higher $M_\bullet/M_\ast$ ratios than numerical mergers, since delayed merger remnants can be slightly overmassive at fixed $M_\ast$.
Secondly, the galaxy merger that precedes the BH merger can drive both $\mathrm{sSFR}$ and BH accretion rate bursts, which boost the XRB and BH luminosity. In \citet{Dong-Paez2023a}, we find that the increased $\mathrm{sSFR}$ has decayed by the time of the delayed merger, and so the galactic XRB boost is absent for delayed mergers. In contrast, the mass gained by BHs during the BH accretion rate boost allows them to accrete at a higher rate on average even at the delayed merger.

An Eddington-limited brightening can greatly increase the observability of BHs in the X-ray. The fraction of both AGN-dominated and observable sources increases significantly with respect to the fiducial accretion rate, to $\sim 45\%-75\%$ ($129$ out of $878$ -- $98$ out of $129$) and $10\%-30\%$ ($98$ out of $878$ -- $36$ out of $129$). We remind the reader that the brightening considered here is not necessarily coincident with the GW chirp, and instead can be produced days to years after the merger when the cavity opened by the binary in the disc is refilled.

In conclusion, a significant fraction of our merger remnants ($5-30\%$) can be detected by high-sensitivity instruments in X-rays and dominate the contaminating galactic emission, especially in the case of a merger-driven brightening. In the UV, galaxies strongly outshine the AGN, rendering almost the whole sample unobservable. We note that here we only consider photometry, but emission lines from the AGN could help disentangle the relative contributions if sufficiently luminous. Since in our model AGN can only be observed in the X-rays, henceforth we restrict our analysis to this band.

\begin{figure*}
    \includegraphics[width=\textwidth]{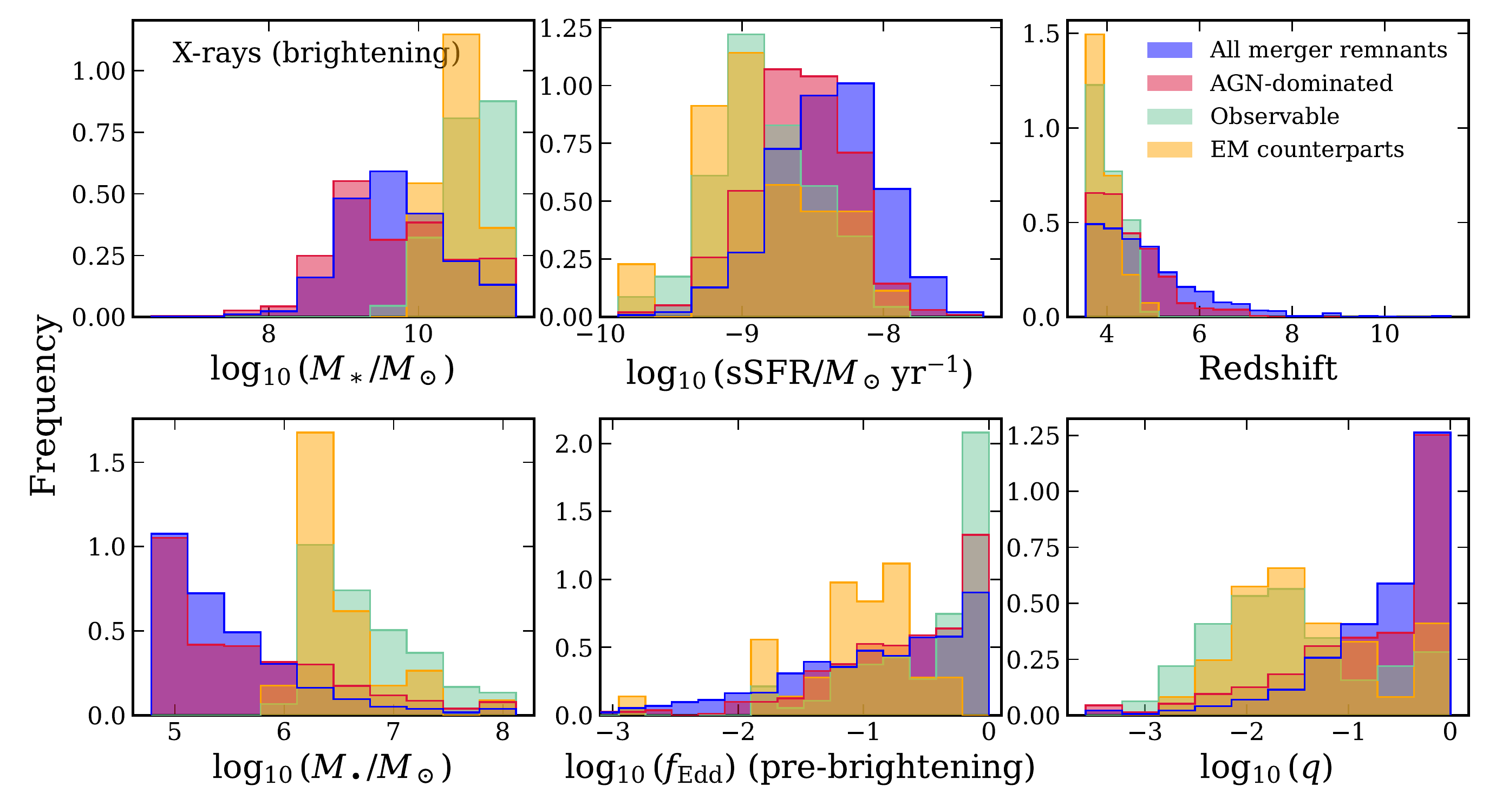}
    \caption{Similar to Fig.~\ref{fig:Xray_pops}, but assuming a brightening luminosity such that $f_\mathrm{Edd}=1$. In orange, we show the distribution of observable mergers according to any of the criteria defined in the text. The presence of a brightening can significantly increase the number of lower-mass X-ray-observable mergers. EM counterparts are biased in a similar way to X-ray-observable remnants at the brightening.}
    \label{fig:Xray_pops_reb}
\end{figure*}

\subsubsection{X-ray transients}
\label{subsubsec:X-ray_transients}

Simply being above the detectability threshold is not a sufficient condition to identify a merger. Mergers also typically require a bright transient signature to ease their identification. In our model, we consider that the accretion rate can attain $f_\mathrm{Edd}=1$ (in a `brightening') around the time of the merger, see Section~\ref{subsec:transients}.

We study the magnitude of merger-driven transient signals in our sample by comparing the pre-transient total flux with the brightening total flux for each source. Here, by total flux we mean the sum of fluxes from the remnant BH and its host galaxy. We recall that we assume that a merger-induced variation in the X-ray flux can be detected if either: (i) The source `appears' -- it is undetected before and detected at the brightening. The BH is observable at the brightening. (ii) The source is detected in both cases, but the flux changes significantly, by more than a factor of $2$. This is probably an optimistic choice since AGN are intrinsically variable sources. Additionally, the BH has to be observable before or after. We denote mergers fulfilling any of these criteria as EM counterparts. 

Transients are explored in Fig.~\ref{fig:Xray_transients}. Pre-transient and brightening fluxes are connected by orange dashed or solid lines if there is an EM counterpart, corresponding observable transients of case (i) or (ii), respectively. We find that $4\%$ of numerical mergers ($37$ out of $878$) have an EM counterpart, according to any of our criteria. For delayed mergers, the fraction rises to almost $20\%$.

As discussed above, a brightening can make observable a large number of AGN that would otherwise be too faint to be observed. A large fraction of EM counterparts ($\sim55\%$) are of type (i) in our notation, i.e. they `appear' at the brightening. We note that many of these events are only marginally observable since our sample is dominated by low-flux sources. These mergers would in practice be hard to detect. Additionally, many of these mergers have very low (pre-transient) BH accretion rates or low mass ratios. The brightening luminosity could be much dimmer than estimated in many of these systems due to the low availability of gas. In this sense, our transient model is optimistic.

Around $45\%$ of the sample is of type (ii), that is, the source is detectable both before and after the merger and the change in the flux at brightening is larger than $2$. This sample is dominated by sources that are initially galaxy-dominated and then become AGN-dominated. In this case, the spectral shape of the source could also change at the brightening.
In contrast, only in $\sim 10\%$ of EM counterparts, the source is AGN-dominated before and at the brightening. 
Remnant BHs which are bright enough to be observable before the brightening, generally already accrete at $f_\mathrm{Edd}\sim 1$. Because of this, the flux difference generated by the brightening is low and such bright mergers are unlikely to have a detectable transient feature.

We note that it is likely that our model overestimates the pre-transient emission and thus underestimates the number of transients of type (ii) since the presence of the cavity can reduce the emission from the disc by more than a factor $\sim 2$ \citep[][and references therein]{Bogdanovic2022}.
In our SED model, if the disc is truncated at $20 GM/c^2$ due to the cavity, the X-ray luminosity of all BHs with spin $\gtrsim0.5$ (which encompasses almost all BHs observable at the brightening) decreases by a factor of $>2$. Assuming a pre-merger decrease in the X-ray luminosity drops by a factor of $2$ would lead instead to a factor of $2$ increase in the number of mergers having a detectable transient. 
It is also possible the merger-induced luminosity burst exceeds the Eddington luminosity \citep{Armitage2002}.

\subsubsection{The population of X-ray-observable mergers}
\label{subsubsec:X-ray_biases}

In the sections above, we identified several sub-samples of BH mergers that are more favourable for X-ray detection. These sub-samples do not reflect the properties of the global population of BH mergers. If future instruments used X-rays to detect BH mergers, they would be biased with respect to the global merger population. In this section, we quantify such biases by studying the differences between AGN-dominated mergers, observable mergers, EM counterparts, and the global population of BH mergers. We will use the numerical merger sample in order to account for the multi-messenger observability of BH mergers since the GW analysis was only performed for this sample. 

Since more massive BHs are generally brighter, observable remnants are strongly biased towards massive mergers with $M_\bullet \gtrsim 10^6\,\Msun$ and $M_\ast \gtrsim 10^{10}\,\Msun$, and sources with a high $M_\bullet/M_\ast$ ratio, which are more likely to be AGN-dominated. This results in AGN-dominated and observable mergers being on average over-massive with respect to the global BH merger population at fixed $M_\ast$, for $10^{9.5}\,\Msun \lesssim M_\ast \lesssim 10^{10.5}\,\Msun$ (Fig.~\ref{fig:MBH_vs_Mgal_Xray}). This effect is akin to the Lauer bias \citep{2007ApJ...670..249L} that causes AGN in a flux-limited sample to yield a relation between BHs and galaxy properties with BH masses above the `true' relation for the full underlying population, especially at high redshift. The exception is observable and AGN-dominated remnants in $M_\ast\gtrsim^{10.5}\,\Msun$ galaxies, which consist of almost all bright high-mass mergers in our sample as they have systematically high accretion rates and $M_\bullet/M_\ast$ ratios and therefore are not biased.

In Fig.~\ref{fig:Xray_pops}, we compare the distributions in $M_\ast$, $M_\bullet$, $\mathrm{sSFR}$, $f_\mathrm{Edd}$, redshift, and $q$ of the global BH merger population, AGN-dominated sources, and observable remnant BHs. 

We first look at the AGN-dominated sources in comparison to the global BH merger population. 
We find that, critically, very high accretion rates are needed for the remnant BH to be bright enough to dominate the galactic emission. Also, BHs hosted by galaxies with low XRB emission (i.e. $\mathrm{sSFR}$) are more likely to be AGN-dominated. This biases the sample of AGN-dominated mergers towards lower $\mathrm{sSFR}$ galaxies at fixed $M_\ast$ and towards high-mass galaxies, which generally have lower $\mathrm{sSFR}$. Finally, high-mass BHs are only assembled at lower redshifts, and so AGN-dominated mergers tend to occur at lower redshifts.  

Observable mergers have similar characteristics, further exacerbated by the flux requirement, which selects only BHs with $M_\bullet \gtrsim 10^6\,\Msun$. This leads to  the selection of host galaxies with $M_\ast \gtrsim 10^{10}\,\Msun$, low $\mathrm{sSFR}$, and $z\lesssim4.5$. Observable remnants are likewise strongly biased towards highly-accreting BHs, with almost all remnants having $f_\mathrm{Edd}>0.3$. The mass ratios are small because numerical BH mergers involving a massive primary often have a much lighter secondary. This effect is much less pronounced for delayed mergers.

In summary, AGN-dominated and observable mergers are strongly biased towards highly-accreting BHs hosted in galaxies with low $\mathrm{sSFR}$. Observable mergers are further biased towards high BH and host galaxy masses.  We note that a significant fraction of the low-$M_\bullet$ AGN-dominated population would not be present if dynamical delays were taken into account since the delay times of BHs in low-mass galaxies are long, so the BHs would grow significantly during the delay or not coalesce before the end of the simulation. 

In Fig.~\ref{fig:Xray_pops_reb}, we consider that the merger produces an Eddington-limited brightening around the time of the merger, increasing the luminosity. In this scenario, a large fraction of sources is AGN-dominated, and so AGN-dominated mergers trace the global merger population well. Observable BHs now include a larger number of lower mass mergers ($M_\bullet \sim 10^6\,\Msun$ and $M_\ast \sim 10^{10}\,\Msun$) that would otherwise not be observable given their low fiducial accretion rates. An important caveat here is that the brightening luminosity might depend on the pre-transient accretion, which our model does not take into account.

The population of EM counterparts is similar to that of observable brightenings. Nevertheless, some of the brightest, most massive BHs are not detected as EM counterparts. This is because high-mass BHs accrete at high rates before the brightening, so the assumed $f_\mathrm{Edd}=1$ brightening does not increase the flux significantly (see Section~\ref{subsubsec:X-ray_transients}). We note that, as discussed in Section~\ref{subsubsec:X-ray_transients}, this population of bright massive mergers could be observable if the presence of the inner disc cavity significantly decreases the pre-transient luminosity.
\begin{figure}
    \includegraphics[width=\columnwidth]{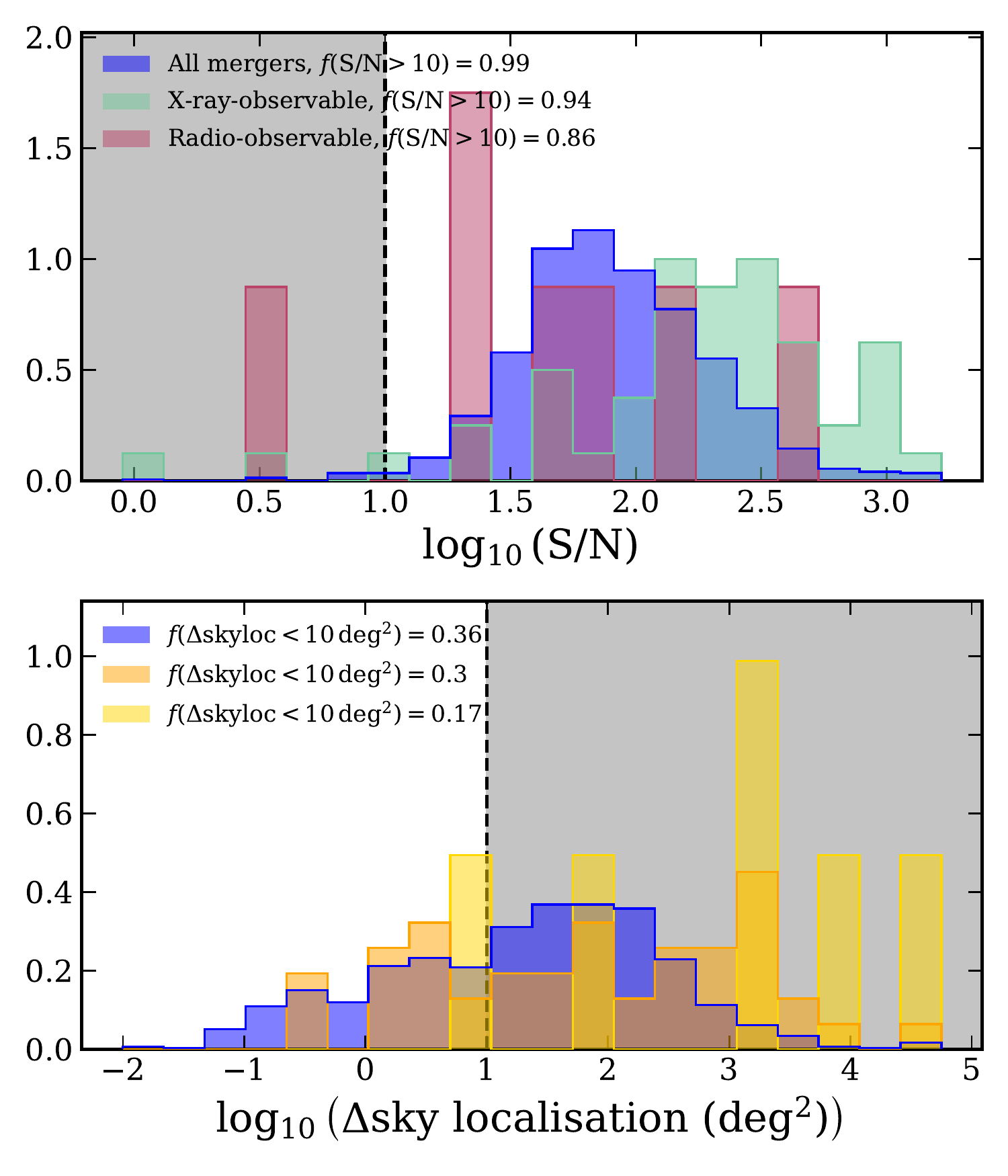}
    \caption{LISA $90\%$-confidence GW sky localisation error of EM-observable mergers. The top panel shows the distribution of all numerical mergers (in blue), X-ray-observable mergers (in green), and radio-observable mergers with SKA in the pessimistic model (in pink). The bottom panel shows the distribution of X-ray EM counterparts (in orange) and radio EM counterparts (in yellow).}
    \label{fig:GW_detectability_EM}
\end{figure}

The population of X-ray-observable mergers can have on average worse GW detectability and parameter estimation than the global merger sample. 
X-ray-observable mergers tend to be biased towards high-mass mergers, which tend to have low mass ratios. Such high-mass, low-$q$ mergers tend to be harder to detect with LISA. In Fig.~\ref{fig:GW_detectability_EM}, we study the GW sky localisation error of the electromagnetically detectable mergers in our numerical merger sample. 
The GW sky localisation of EM counterparts is similar to that of the global numerical merger sample. $31\%$ of X-ray EM counterparts are localised with $90\%$ confidence accuracy smaller than $10\,\mathrm{deg}^2$, which is comparable with the value of $37\%$ for all mergers. This fraction is lower ($15\%$, $6$ out of $41$) for X-ray-observable remnants tends since they are more strongly biased towards high-mass low-$q$ mergers which are poorly detected. 
We find that most X-ray-bright mergers can in general also be detected by LISA, with a $\mathrm{S/R}$ distribution comparable to that of the global merger population. Of the $46$ X-ray-observable mergers, $41$ are also GW-detected.  

In conclusion, the population of AGN-dominated sources, observable sources, and EM counterparts are biased tracers of the underlying merger population. Observable BHs and mergers are more massive, inhabit more massive and less star-forming galaxies, accrete at higher rates, and occur at lower redshifts. They also tend to be overmassive with respect to the $M_\bullet$-$M_\ast$ relation. 

Finally, we remind the reader that our discussion is limited to high-$z$ events since the simulation stops at $z\sim 3.5$. At lower redshift, a larger fraction of the merger population could be detectable since the sources will be closer to the observer. Furthermore, we have not included delayed mergers here, whose lower redshift and mass ratios closer to unity would improve the S/R and sky localisation.

\subsection{Radio}
\label{subsec:radio}

\subsubsection{Radio detectability}
\label{subsubsec:radio_detectability}
\begin{figure*}
    \centering
    \includegraphics[width=\textwidth]{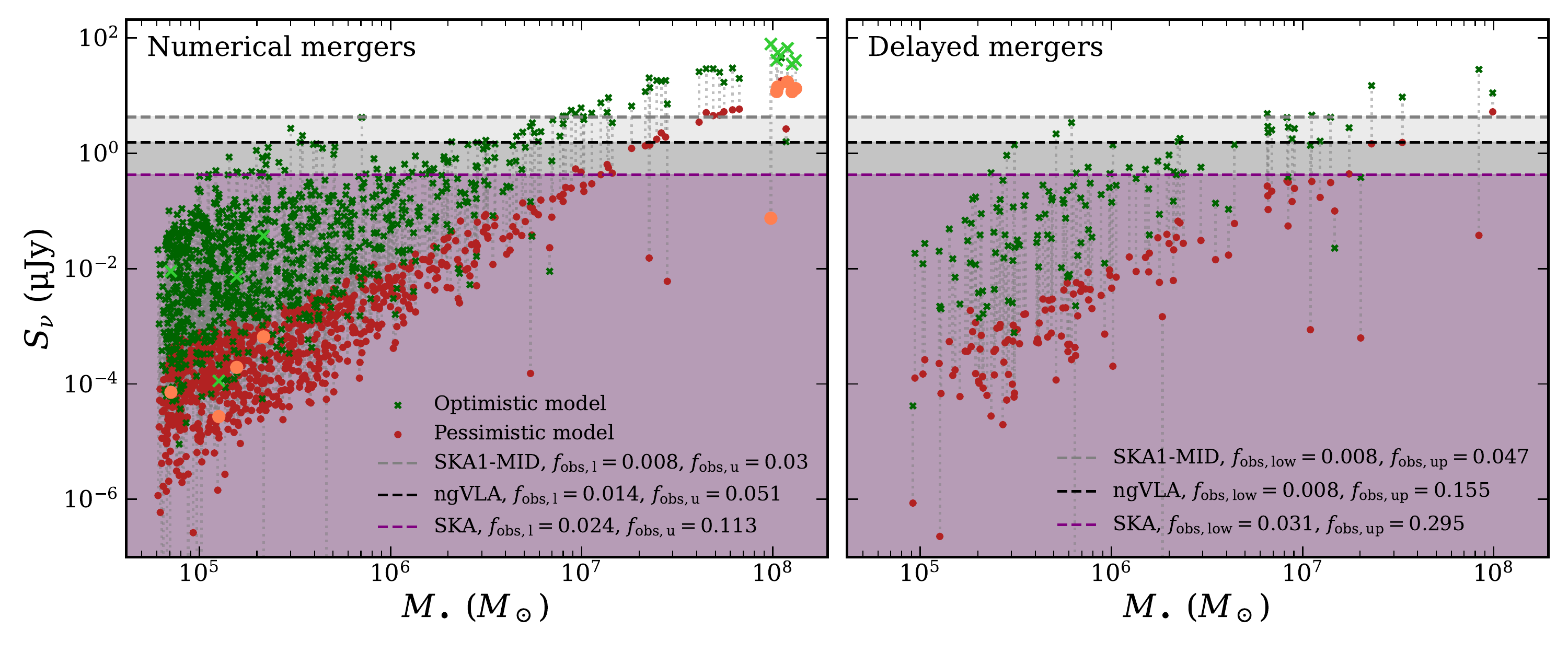}
    \caption{Observer-frame spectral flux density at $\nu=\SI{2}{\giga\hertz}$ of numerical merger remnants against the remnant BH mass, for numerical mergers (left panel) and delayed mergers (right panel). For each sample, the `lower limit' model \citep[based on the fundamental plane,][]{Gultekin2009} is shown with dark red circles, and the `upper limit' model \citep[based on the theoretical model in][]{Meier2001} is shown with dark green crosses. The dashed horizontal lines represent the $5\sigma$ sensitivity thresholds for the ngVLA \citep{Carilli2015}, in black, for the SKA1-MID, in grey, and for the SKA \citep{Prandoni2015}, in purple. These thresholds are calculated assuming \SI{9}{hr} exposure times and that the observed sources are not resolved. The legend indicates the fraction of observable events, which lie above the sensitivity threshold of each instrument and are observable with LISA, if applicable. Mergers that are undetected with LISA are shown with bigger and lighter markers.}
    \label{fig:radio_jet_observability}
\end{figure*}

The radio observability of merger remnants jets at $\nu=\SI{2}{\giga\hertz}$ (observer-frame) is explored in Fig.~\ref{fig:radio_jet_observability}. We recall that we consider two models: a lower limit model for the core radio emission based on the fundamental plane \citep[following][]{Gultekin2009}, and an upper limit model for the total radio luminosity based on the theoretical model in \cite{Meier2001}. We also recall that we do not explicitly consider the contamination due to the galactic radio emission in our analysis since we cannot quantify this reliably (see Section~\ref{subsec:galactic_emission}). Some of the sources which exceed the instrumental sensitivity threshold could be outshone by their hosts if $M_\bullet\lesssim 10^{7.5} M_\odot$. In this section, we disregard this effect and denote remnant BHs as observable if their flux exceeds the instrumental sensitivity and for numerical mergers if they are also detected by LISA.

We consider the detectability by future surveys with ngVLA for which we consider a sensitivity threshold $1.5\,\si{\micro\jansky}$ \citep{Carilli2015} and SKA1-MID with a threshold of $4.2\,\si{\micro\jansky}$ \citep{Prandoni2015}. The thresholds are calculated assuming \SI{9}{hr} exposure observations at $2\,\si{\giga\hertz}$ and that sources are not resolved. We also consider that the full SKA could go deeper, and assume a sensitivity of $0.4\,\si{\micro\jansky}$. These detectability limits are denoted by grey, black, and purple dashed lines in Fig. \ref{fig:radio_jet_observability} respectively.

We find that a fraction of mergers can be detected in the radio with future instruments, although this fraction depends strongly on the model assumed for the radio luminosity and on the instrument's sensitivity. For the pessimistic empirical model for core emission, only the most massive BHs with masses $M_\bullet \gtrsim 10^{7.5}\,\Msun$ can be detected. The fraction of observable mergers is in the range $1\%-3\%$, which is lower than the fraction we found for X-rays in Section~\ref{subsubsec:UV_X-rays_detectability}. For the optimistic theoretical model for the radio emission, BHs with mass $M_\bullet \gtrsim 10^{6.5}\,\Msun$ are generally above the flux limit. The fraction of observable mergers rises significantly to $4\%-30\%$, consistent with the fractions found for the X-rays detectability. 
The different predictions for our two models stem from the fact that the pessimistic model estimates only the core emission while our optimistic model estimates the total emission, although it is also possible that the pessimistic model underestimates the radio luminosity for highly accreting AGN \citep{Gultekin2022}.

\subsubsection{Radio transients}
\label{subsubsec:radio_transients}

In radio, we consider two possible types of transients. Firstly, analogous to our model for X-ray transients, we consider a merger brightening caused by a sharp increase in the accretion rate. Secondly, we consider a flare, as found in some simulations (see Section~\ref{subsec:transients} for more details). 
As for the X-ray transients, we consider three possibilities for detecting an EM counterpart, (i) The source is detected only after the merger, (ii) The source is detected before and after, but the flux changes by more than a factor of $2$. As in the previous section, we do not take into consideration the galactic contribution to the flux. Henceforth, we take the fundamental plane pessimistic model as the fiducial radio model for this analysis since it models only the core luminosity, which is expected to show stronger flux variations on timescales of weeks to months, and we optimistically take SKA as the fiducial instrument. Since the GW analysis was only performed on numerical mergers, we use numerical mergers as the reference sample.

\begin{figure}
    \includegraphics[width=\columnwidth]{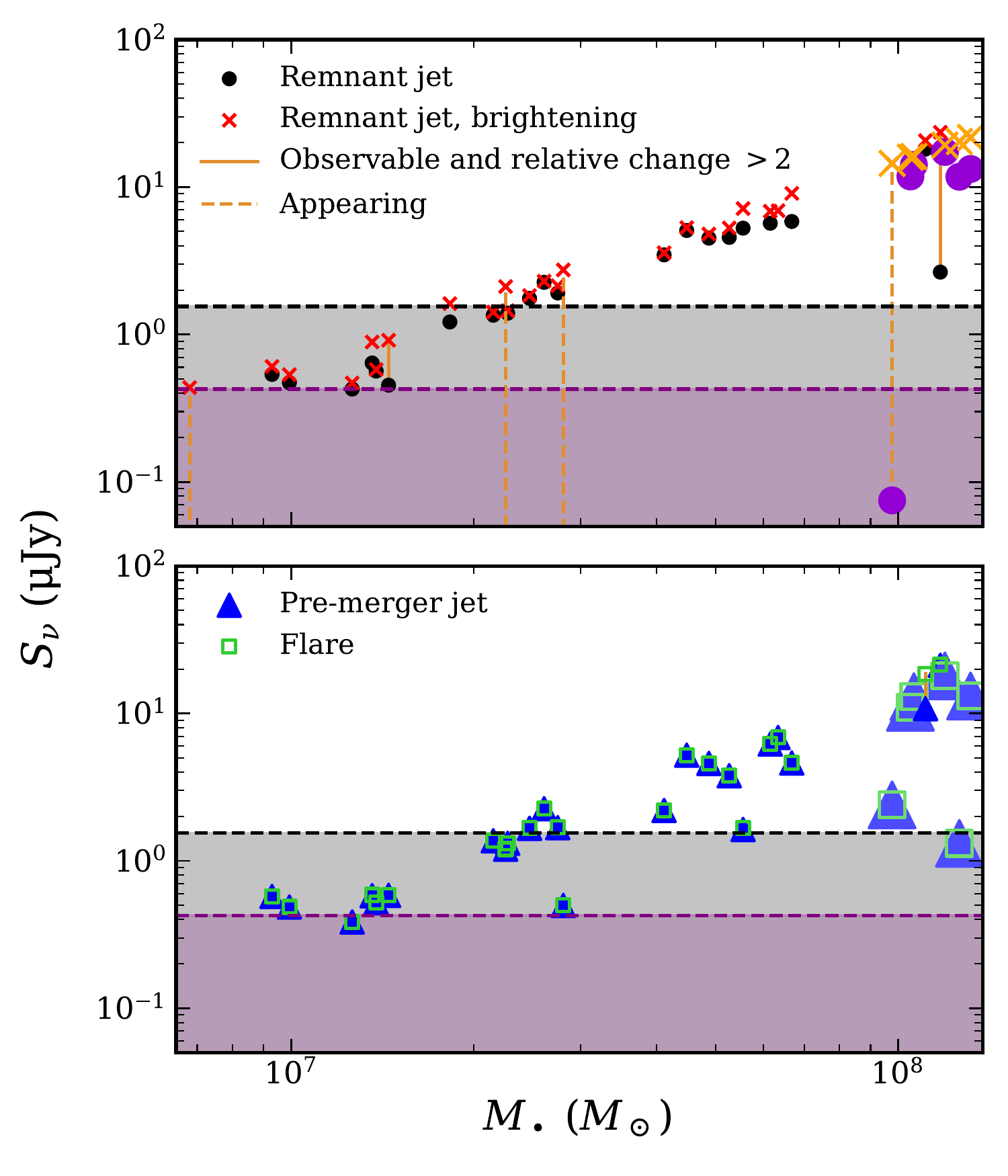}
    \caption{Observability of a transient generated by a $f_\mathrm{Edd}=1$ brightening (top panel) or a merger-induced flare (bottom panel) in the numerical merger sample. Black dots and red crosses correspond to the post-merger BH radio fluxes for the fiducial and the rebrightening scenario, respectively. The pre-merger primary BH radio flux is shown in blue triangles, and the flare flux is shown in green squares. We only show mergers for which either the fiducial or the merger-enhanced (brightening or flare) flux is observable. Fiducial and merger-enhanced fluxes are connected by orange dashed lines if the fiducial flux is undetected and the enhanced flux is detected, or orange solid lines if the flux is detected in both cases and the flux change is larger than a factor of $2$. For simplicity, we only consider the detectability of the pessimistic model by the SKA (assuming a sensitivity of $0.4\,\si{\micro\jansky}$, purple horizontal line), but the ngVLA sensitivity is also shown for reference ($1.5\,\si{\micro\jansky}$, black horizontal line). GW-undetected events are shown with bigger, brighter markers.
    With our modelling, most of the merger-induced flux changes are too small to be detectable in radio.}
    \label{fig:radio_jet_transients}
\end{figure}

In Fig.~\ref{fig:radio_jet_transients} we show results for the brightening (top panel) and the flare (bottom panel). We find that observing a radio EM counterpart is unlikely -- less than $1\%$ of mergers have a detectable merger-induced transient. 
$\sim 80\%$ of EM counterparts are observable due to a post-merger brightening, i.e. an increase in the accretion rate to $f_\mathrm{Edd}=1$. Most potentially observable brightenings already accrete at very high rates before the merger and so the merger-induced change in luminosity is small, which means such mergers will not be recognisable as radio transients. Transients EM counterparts are rarer in the radio than in the X-ray as at fixed BH mass $L_\mathrm{R,aft}/L_\mathrm{R} = (L_\mathrm{X,aft}/L_\mathrm{X})^{0.67}$ (eq.~\ref{eq:FP}), where the suffix `aft' denotes the brightening luminosity, the flux variation due to the brightening in radio is smaller than in X-rays. 
Only one EM counterpart ($\sim 20\%$ of all EM counterparts, bottom panel of Fig. \ref{fig:radio_jet_transients}) is observable due to a flare. The amplitude of the flare is strongly dependent on the mass ratio, and thus only very bright and massive mergers with nearly equal mass ratios, which are very uncommon in our sample, can have an observable flare. In general, the number of EM counterparts is small also because the number of observable BHs is small.

Finally, we note that these estimates using the pessimistic model based on the fundamental plane are likely lower limits since the fundamental plane can underestimate the radio luminosity of highly accreting BHs \citep{Gultekin2022}.

\subsubsection{The population of radio-observable mergers}
\label{subsubsec:radio_biases}

\begin{figure*}
    \includegraphics[width=\textwidth]{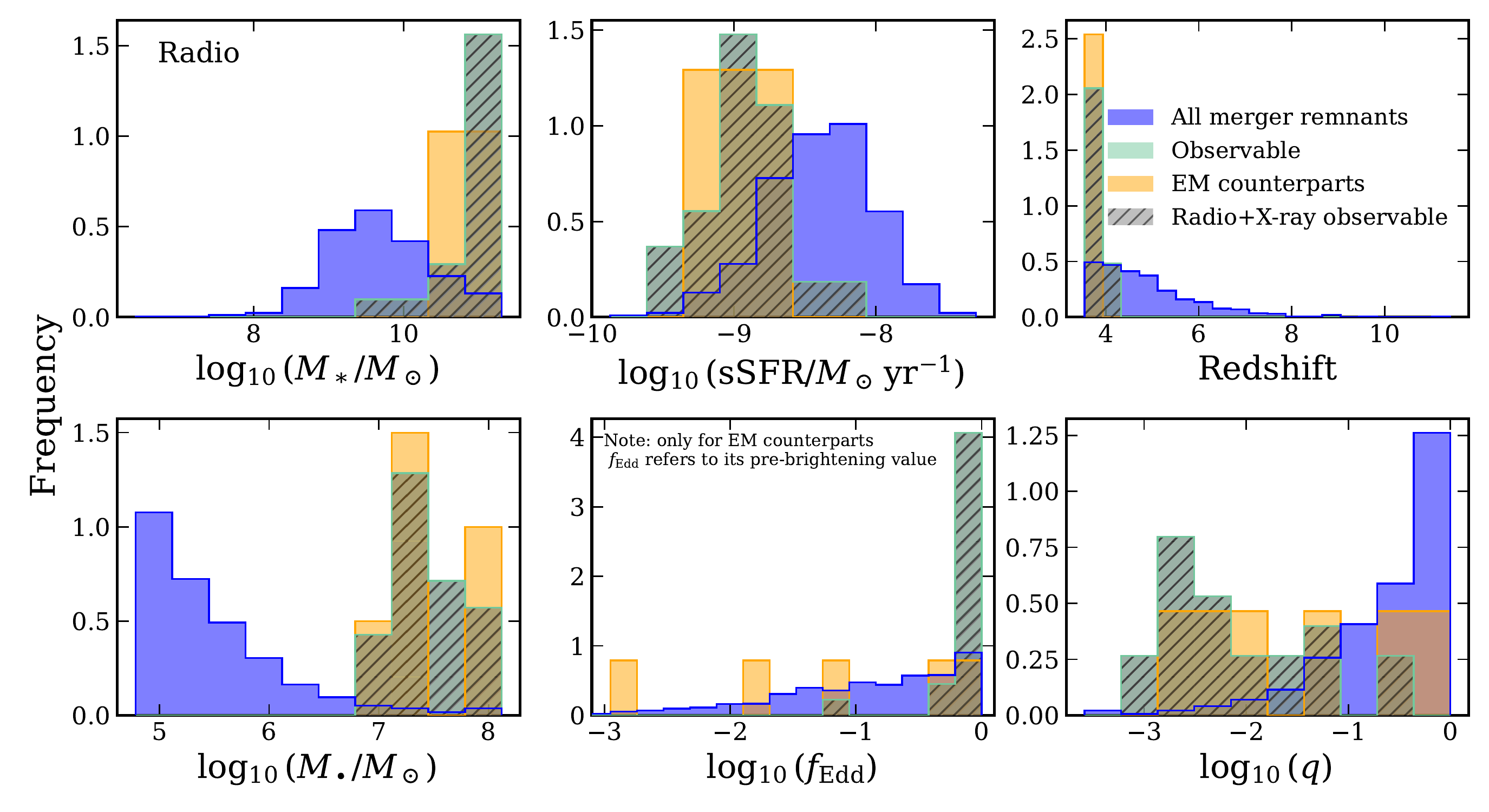}
    \caption{Distribution of $M_\ast$, $M_\bullet$, $\mathrm{sSFR}$, $f_\mathrm{Edd}$, redshift, and $q$ for the overall population of numerical BH mergers (in blue), for radio-observable mergers (AGN-dominated mergers where $F_{\mathrm{X},\bullet}>F_{\mathrm{X,lim}}$, in green) and for EM counterparts (in orange). All histograms are normalised to unity. Observable BH mergers are hosted in more massive galaxies, have lower $\mathrm{sSFR}$ and higher $f_\mathrm{Edd}$ with respect to the general BH-merging population. All radio-observable mergers are also X-ray-observable.}
    \label{fig:radio_pops}
\end{figure*}
The population of radio-observable remnant BHs is also a biased tracer of the underlying remnant BH population. In order to quantify this, in Fig.~\ref{fig:radio_pops} we compare the properties of observable mergers and mergers with transient EM counterparts to the global BH merger population. We also show the population of radio-observable mergers which are also X-ray-observable. Again we use the pessimistic model and SKA as the fiducial model and instrument. For a total of $878$ numerical mergers, there are $21$ radio-observable mergers and $5$ EM counterparts. The biases of the radio-observable population are qualitatively identical to those of the X-ray observable population -- observable BHs have on average significantly higher $M_\bullet$, $M_\ast$ and $f_\mathrm{Edd}$ and lower $\mathrm{sSFR}$ and redshift. However, given that a smaller amount of mergers are observable for our pessimistic model, the sample is more strongly biased towards high BH and galaxy masses ($M_\bullet\gtrsim 10^7\,\Msun$, $M_\ast\gtrsim 10^{10}\,\Msun$) and low redshifts ($z\lesssim 4$). Radio-observable mergers are also overmassive with respect to their galaxies. 
We find that all radio-observable mergers in our sample are also X-ray-observable since X-ray bright BHs tend to also be bright in the radio.

As discussed in the previous section, the number of EM counterparts is very small since neither the brightening nor the flare is able to sufficiently change the flux for the brightest BHs and produce an observable transient. The population of EM counterparts is similar to that of observable mergers, although the number of these mergers is too low to carry out a meaningful statistical analysis.

Since radio-observable mergers are strongly biased towards high-mass low-$q$ mergers in our pessimistic model, they have poorer GW parameter estimation with LISA. As shown in Fig~\ref{fig:GW_detectability_EM}, few radio-observable remnants or radio EM counterparts have a $90\%$-confidence error lower than $10\,\mathrm{deg}^2$. Despite this, radio-observable remnants and radio EM counterparts tend to be detectable with LISA, although GW-undetected mergers constitute a fraction of $\sim20\%$ ($5$ out of $26$ radio-observable mergers). Finally, we refer the reader again to the note at the end of Section~\ref{subsubsec:X-ray_biases}, which applies also to radio sources. 
As before, we note that the results presented in this section are likely lower limits to the radio observability of merger remnants since our pessimistic model is likewise a lower limit on the luminosity of highly accreting BHs \citep{Gultekin2022}.

\section{Comparison with previous work}
\label{sec:comparison}

Previous work by \cite{Tamanini2016} and \citet{Mangiagli2022} also studied the possibility of joint (EM and GW) multi-messenger detections of BH mergers. They use a BH population synthesised from a semi-analytic model of galaxy formation.
The X-ray AGN fluxes estimated by \citet{Mangiagli2022} are similar to those presented in this work. The fraction of detected radio sources found in the present work is significantly lower than the values reported in \citet{Tamanini2016} and \citet{Mangiagli2022}, despite the optimistic model presented here being conceptually similar. Our luminosities can be more than $4$ orders of magnitude smaller, for two reasons: Firstly, for our optimistic model we assume a conversion factor from jet power to radio luminosity of $10^{-2}$, instead of assuming full or very efficient conversion, and take into account the fact that the radiation is distributed across a wide synchrotron spectrum. 
Secondly, we use a different model for the flare emission, which predicts smaller luminosities. \citet{2023MNRAS.519.5962L} used similar methods as \citet{Tamanini2016} and \citet{Mangiagli2022} for the synthesis but focus on the GW localisation of BH mergers, by studying the galaxy fields in LISA error-boxes.

Other previous studies \citep[e.g.][]{2019MNRAS.485.1579K,2019ApJ...879..110K} have rather focused on BH binary signatures, such as a periodic modulation of the BH luminosity or spectral features. Many of these signatures are expected already well before the BH coalescence and in only few cases they can be observed concurrently with a GW detection with LISA, although they can provide valuable information to break degeneracies and compare the speed of photons and gravitons \citep{2017PhRvD..96b3004H}.

The main novelty of the present study with respect to previous work is that we use a hydrodynamical simulation that follows self-consistently the evolution of BHs and their environments and obtain a realistic BH merger population. This also allows us to calculate consistently environmental parameters such as gas and dust obscuration. We also introduce physically motivated models for the AGN SED, radio emission, and subgrid obscuration. Further, we study several important observational effects: (i) the contamination from the host galaxy emission (ii) the observability of a possible merger transient signal occurring around the time of the merger (iii) the observational biases of the EM-observable population.

\section{Conclusions}
\label{sec:conclusions}
In this work, we have presented a comprehensive study of BH mergers in \textsc{Obelisk}, a cosmological hydrodynamical simulation following the evolution of a protocluster down to redshift $z \sim 3.5$. Building on \citet{Dong-Paez2023a}, which studied the properties of the population of BH mergers and compared it with the underlying global population of main BHs, we have performed a multi-messenger analysis of the detectability of BH mergers, in order to forecast their detectability and assess the possible observational biases. We performed a GW and EM analysis of numerical BH mergers at the resolution of the simulation and an EM analysis of delayed BH mergers which consider post-processed dynamical delays below the simulation resolution. 
We summarise our results below: 
 \begin{itemize}
    \item Most of the numerical merger sample ($\sim 99\%$) can be detected by LISA, generally with very high S/R (Fig.~\ref{fig:GW_analysis}). Only a small sample of high-mass low-$q$ mergers and low-mass high-$z$ mergers are undetected. The intrinsic binary parameters, such as the BH masses, spins, and redshift, can generally be measured with high accuracy. Only $\sim 37\%$ of these high-z BH mergers can be localised in the sky with $90\%$-confidence error better than $10\,\mathrm{deg}^2$.
    
    \item In UV, remnant BHs are significantly fainter than their host galaxies, which are actively star-forming at the redshifts considered ($z>3.5$).

    \item In X-rays, $5\%-15\%$ of remnant BHs are bright enough to be detectable by future instruments while dominating over their host galaxy's emission (Fig.~\ref{fig:L_BH_vs_L_gal}). If a merger-induced brightening increases the BH accretion rate to the Eddington rate, up to $10\%-30\%$ could become observable and in some cases be identifiable as a transient EM counterpart (Fig.~\ref{fig:Xray_transients}).
    
    \item X-ray observable BH remnants tend to accrete near the Eddington limit, have higher $M_\ast$ ($\gtrsim 10^{10}\,\Msun$), and $M_\bullet$ ($\gtrsim 10^6\,\Msun$), occur at lower redshift, in lower $\mathrm{sSFR}$ galaxies (Figs.~\ref{fig:Xray_pops} and~\ref{fig:Xray_pops_reb}), and are overmassive at fixed galaxy mass with respect to the global population for $M_\ast \lesssim 10^{10.5}\,\Msun$ (Fig.~\ref{fig:MBH_vs_Mgal_Xray})

    \item In radio between $\sim 1\%$ and $30\%$ of mergers can be detected by future instruments (Fig.~\ref{fig:radio_jet_observability}). A merger-induced increase of the core radio luminosity is found to lead to only a small number of transients detectable as EM counterparts (Fig.~\ref{fig:radio_jet_transients}). 
    
\item The population of radio-observable BH merger remnants differ with respect to the full merger population. The biases with respect to the full merger population are qualitatively analogous to the X-ray observable sample (Fig.~\ref{fig:radio_pops}). Most radio-observable mergers are also X-ray observable.

\end{itemize}

Overall, we found that the number of EM counterparts is currently limited by LISA's ability to localise the systems in the sky. It is worth noting that an additional LISA-like detector would dramatically change the situation, leading to a sky localisation improvement of two orders of magnitude \citep{2020NatAs...4..108R} in which case we would instead be limited by the sensitivity of EM telescopes and by the presence or absence of a tell-tale transient EM sign of BH mergers.

The expectation is that many if not most LISA BH mergers will have $z \gtrsim 2$, therefore our work considers some of the most plausible types of sources, but high-redshift sources are by definition fainter than low-redshift sources at fixed luminosity. At lower redshifts, prospects for looking for EM counterparts are brighter \citep{2023MNRAS.519.5962L}. In future work, we will explore the detectability of BH mergers in \textsc{Obelisk} at additional wavelengths.

\begin{acknowledgements}
We thank Geoffrey Bicknell, Sera Markoff, Stanislav Babak, and Alexander Wagner for stimulating discussions, and Sylvain Marsat for allowing usage of the \texttt{lisabeta} code for this paper. We thank the anonymous referee whose comments and suggestions helped improve the manuscript. MV, YD, NW, and SV acknowledge funding from the French National Research Agency (grant ANR-21-CE31-0026, project MBH\_waves). MT acknowledges support from the NWO grant 0.16.VIDI.189.162 (`ODIN'). AM acknowledges support from the postdoctoral fellowships of IN2P3 (CNRS). This project has received funding from the European Union’s Horizon 2020 research and innovation programme
under the Marie Skłodowska-Curie grant agreement No. 101066346 (MASSIVEBAYES). This work has received funding from the Centre National d’Etudes Spatiales.
This work has made use of the Horizon Cluster hosted by Institut d’Astrophysique de Paris; we thank Stéphane Rouberol for running smoothly this cluster for us. We acknowledge PRACE for awarding us access to Joliot Curie at GENCI@CEA, France, which was used to run most of the simulations presented in this work. Numerical computations were partly performed on the DANTE platform, APC, France. Additionally, this work was granted access to the HPC resources of CINES under allocations A0040406955 and A0040407637 made by GENCI.

\end{acknowledgements}


\bibliography{references}

\end{document}